\documentclass[sigconf]{acmart}

\author{Jeremiah Blocki}
\affiliation{%
	\institution{Purdue University}
}
	\email{jblocki@purdue.edu}
\author{Wuwei Zhang}
\affiliation{%
	\institution{Purdue University}
}
	\email{zhan1015@purdue.edu}

\usepackage{balance,color,amsmath, alltt, xspace, epsfig, algorithm, subfigure, xcolor, multirow, psfrag,mathtools,comment, verbatim,algpseudocode,grffile}

\newcommand{\ignore}[1]{}

\newcommand{\mypara}[1]{\vspace*{0.05in}\noindent\textbf{#1} \xspace}		

\newcommand{\PasswordOfU}[1]{pw_{#1}}

\newcommand{\KOfU}{K_u}

\newcommand{\AllUser}{\mathcal{U}}
\newcommand{\RankRPassword}[1]{pw_{#1}}

\newcommand{\Lap}{\mathsf{Lap}\xspace}								
\newcommand{\Adversary}{\ensuremath{\mathcal{A}}}							

\newcommand{\ZX}{\ZXCVBN} 			

\newcommand{\ZXCVBN}{\mathsf{ZXCVBN}}
\newcommand{\Estimator}{\mathsf{Estimator}}
\newcommand{\EstF}[1]{\textsf{Estimate}(#1)}
\newcommand{\EstimateF}[2]{\textsf{Estimate}(#1,#2)}			
\newcommand{\EstP}[1]{\textsf{p}(#1)}
\newcommand{\EstimateP}[2]{\textsf{p}(#1,#2)}			
\newcommand{\FMPPF}{\ensuremath{\mathsf{FMPPF}}}							
\newcommand{\DAB}{\ensuremath{\mathsf{DAB}}}	
\newcommand{\PK}{\ensuremath{\mathsf{PK}}}							
\newcommand{\SampledData}[1]{\mathcal{D}_{{#1}}}
\renewcommand{\Pr}[1]{\ensuremath{\mathsf{Pr} \left[#1\right] }\xspace}			
\newcommand{\KPsiDALock}[2]{\ensuremath{(#1, #2)\text{-}\mathsf{DALock}}\xspace}		
\newcommand{\hitCountThreshold}{\Psi}
\newcommand{\hitCountThresholdOfU}[1]{\hitCountThreshold_{#1}}	
\newcommand{\DP}[2]{\textsf{DP}(#1,#2)}
\newcommand{\strikeThreshold}{K}	
\newcommand{\strikeThresholdOfU}[1]{\strikeThreshold_{#1}}
\newcommand{\DALock}{\mathsf{DALock}\xspace}	
\newcommand{\AllPassword}{\mathcal{P}}
\newcommand{\CountSketch}{\mathsf{CS}}
\newcommand{\CountSketchCounter}{\mathsf{CS.T}}

\newcommand{\TrueP}[1]{\ensuremath{\mathsf{P}\left(#1\right)}}
\newcommand{\TrueF}[1]{\ensuremath{\mathsf{F}\left(#1\right)}}
\newcommand{\TrueFInD}[2]{\ensuremath{\mathsf{F}\left(#1, #2\right)}}

\newcommand{\Add}[2]{\ensuremath{\mathsf{Add}(#1,#2)}}

\newcommand{\NP}{\mathsf{NP}\xspace}

\newcommand{\FPTAS}{\mathsf{FPTAS}\xspace}
\newcommand{\PTAS}{\mathsf{PTAS}\xspace}

\algrenewcommand\algorithmicindent{1.0em}%
\usepackage{tikz}
\usetikzlibrary{matrix,calc,shapes}
\tikzset{
  treenode/.style = {shape=rectangle, rounded corners,
                     draw, anchor=center,
                     text width=5em, align=center,
                     top color=white, bottom color=blue!20,
                     inner sep=1ex},
  decision/.style = {treenode, diamond, inner sep=0pt},
  root/.style     = {treenode, font=\Large, bottom color=red!30},
  env/.style      = {treenode, font=\ttfamily\normalsize},
  finish/.style   = {root, bottom color=green!40},
  dummy/.style    = {circle,draw}
}

\setcopyright{none} 
\acmYear{2020}

\begin{document}
\title{$\DALock$:  \underline{D}istribution \underline{A}ware Password Throttling} 

\begin{abstract}

Large-scale online password guessing attacks are wide-spread and continuously qualified as one of the top cyber-security risks. The common method for mitigating the risk of online cracking is to lock out the user after a fixed number ($K$) of consecutive incorrect login attempts. Selecting the value of $K$ induces a classic security-usability trade-off. When $K$ is too large a hacker can (quickly) break into a significant fraction of user accounts, but when $K$ is too low we will start to annoy honest users by locking them out after a few mistakes. Motivated by the observation that honest user mistakes typically look quite different than the password guesses of an online attacker, we introduce $\DALock$ a {\em distribution aware} password lockout mechanism to reduce user annoyance while minimizing user risk. As the name suggests, $\DALock$ is designed to be aware of the frequency and popularity of the password used for login attacks while standard throttling mechanisms (e.g., $K$-strikes) are oblivious to the password distribution. In particular, $\DALock$ maintains an extra ``hit count" in addition to ``strike count" for each user which is based on (estimates of) the cumulative probability of {\em all} login attempts for that particular account. We empirically evaluate $\DALock$ with an extensive battery of simulations using real world password datasets. In comparison with the traditional $K$-strikes mechanism we find that $\DALock$ offers a superior security/usability trade-off. For example, in one of our simulations we are able to  reduce the success rate of an attacker to $0.05\%$ (compared to $1\%$ for the $10$-strikes mechanism) whilst simultaneously reducing the unwanted lockout rate for accounts that are not under attack to just $0.08\%$ (compared to $4\%$ for the $3$-strikes mechanism).  
\end{abstract}


\maketitle


 	\section{Introduction}\label{sec: Introduction}
	An online password attacker repeatedly attempts to login to an authentication server submitting a different guess for the target user's password on each attempt. The human tendency to pick weak (``low-entropy'') passwords has been well documented e.g., ~\cite{SP:Bonneau12}. An untargetted online attacker will typically submit the most popular password choices consistent with the password requirements (e.g., ``Password1''), while a targetted attacker~\cite{CCS:WZWYH16} might additionally incorporate background knowledge about the specific target user (e.g., birthdate, phone number, anniversary etc...). To protect user's against online attackers most authentication servers incorporate some form of throttling mechanism. In particular, the $\strikeThreshold$-strikes mechanism temporarily locks a user's account if $\strikeThreshold$-consecutive incorrect passwords are attempted within a predefined time period (e.g., $24$ hours). Setting the lockout parameter $\strikeThreshold$ induces a classic security-usability trade-off. Selecting small values of $\strikeThreshold$ (e.g., $\strikeThreshold=3$) provides better protection against online attackers, but may result in many unwanted lockouts when an honest user miss-types (or miss-remembers) their password. Selecting a larger value of $\strikeThreshold$ (e.g., $\strikeThreshold=10$) will reduce the unwanted lockout rate, but may increase vulnerability to online attacks. 
	
Bonneau et al.~\cite{SP:BHVS12} considered many proposed replacements for password authentication finding that all proposals have some drawbacks compared when compared with passwords. For example, passwords are easy to revoke unlike biometrics. Similarly, hardware tokens are expensive to deploy and require users to carry them around. By contrast, passwords are easy to deploy and do not require users to carry anything around. Put simply we have not found a ``silver bullet'' replacement for passwords. Thus, despite all of their short-comings (and many attempts to replace them) passwords will likely remain entrenched as the dominant form of authentication on the internet~\cite{PasswordPersistence}. Thus, protecting passwords against online attacks without locking out legitimate users remains a crucial challenge for the foreseeable future~\cite{DuoWeakPassword,DictionaryAttack:Ransomware,DictionaryAttack:Microsoft}. 

One approach to protect user's against online guessing attacks is to adopt strict password composition policies to prevent user's from selecting weak passwords. However, it has been well documented that users dislike restrictive policies and often respond in predictable ways~\cite{KSKMBCCE:SIGCHI11}. Another defense is to store cookies on the user's device to prove that the next login attempt came from a known device. Similarly, one can also utilize features such as IP address, geographical location, device and time of day~\cite{sandhu2005system,gordon2014efficiently,NDSS:FJDBG16} to help distinguish between malicious and benign login attempts. While these features can be helpful indicators they are not failproof. Honest users oftentimes travel and login from different devices at unusual times. Similarly, an attacker may attempt to mimic login patterns of legitimate users e.g., using a botnet the online attacker can submit guesses from a vide variety of IP addresses and geographical locations. 
	
\vspace{-0.1cm}
	\subsection{Contributions} 
	We introduce $\DALock$, a novel \underline{D}istribution \underline{A}ware throttling mechanism which can achieve a better balance between usability and security. The key intuition behind $\DALock$ is to base lockout decisions on the {\em popularity} of the passwords that are being guessed. An online attacker will typically want to attempt to login with the most popular passwords  to maximize his chances of success. By contrast, when an honest user miss types (or miss remembers) his password he will typically not be globally popular password. In addition to keeping track of $\strikeThresholdOfU{u}$ (the number of consecutive incorrect login attempts), $\DALock$ keeps track of a ``hit count'' $\hitCountThresholdOfU{u}$ for each user $u$, where $\hitCountThresholdOfU{u}$ intuitively represents the cumulative probability mass of all incorrect login attempts for user $u$'s account. When $\hitCountThresholdOfU{u}$ exceeds the threshold $\hitCountThreshold$ we decide to lock the account. 
	
\begin{figure}[htb]
		\begin{center}
			\includegraphics[height=2in,width=\linewidth]{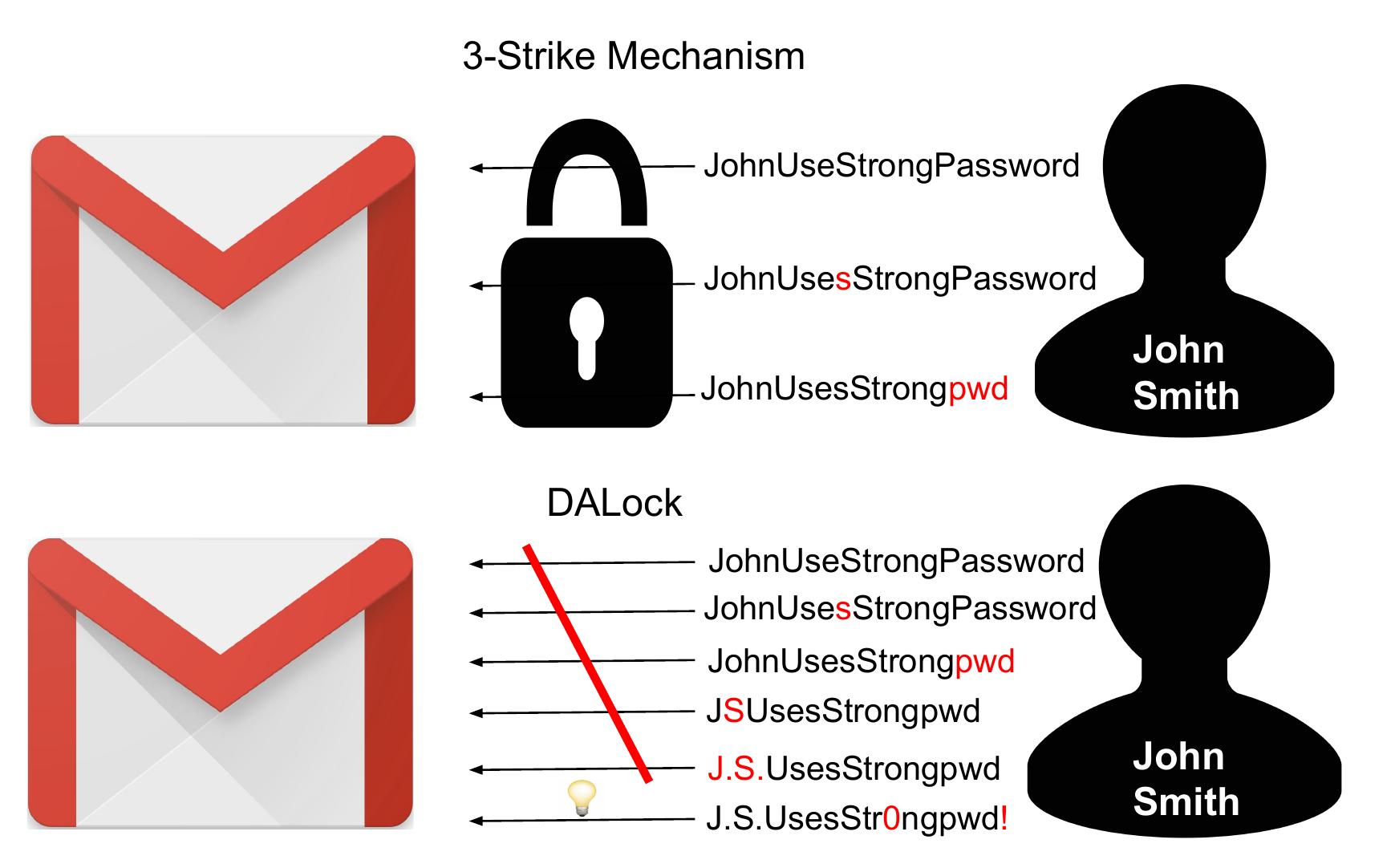}
			\caption{Usability Comparison}\label{figure:introduction_usability}
		\includegraphics[height=2in,width=\linewidth]{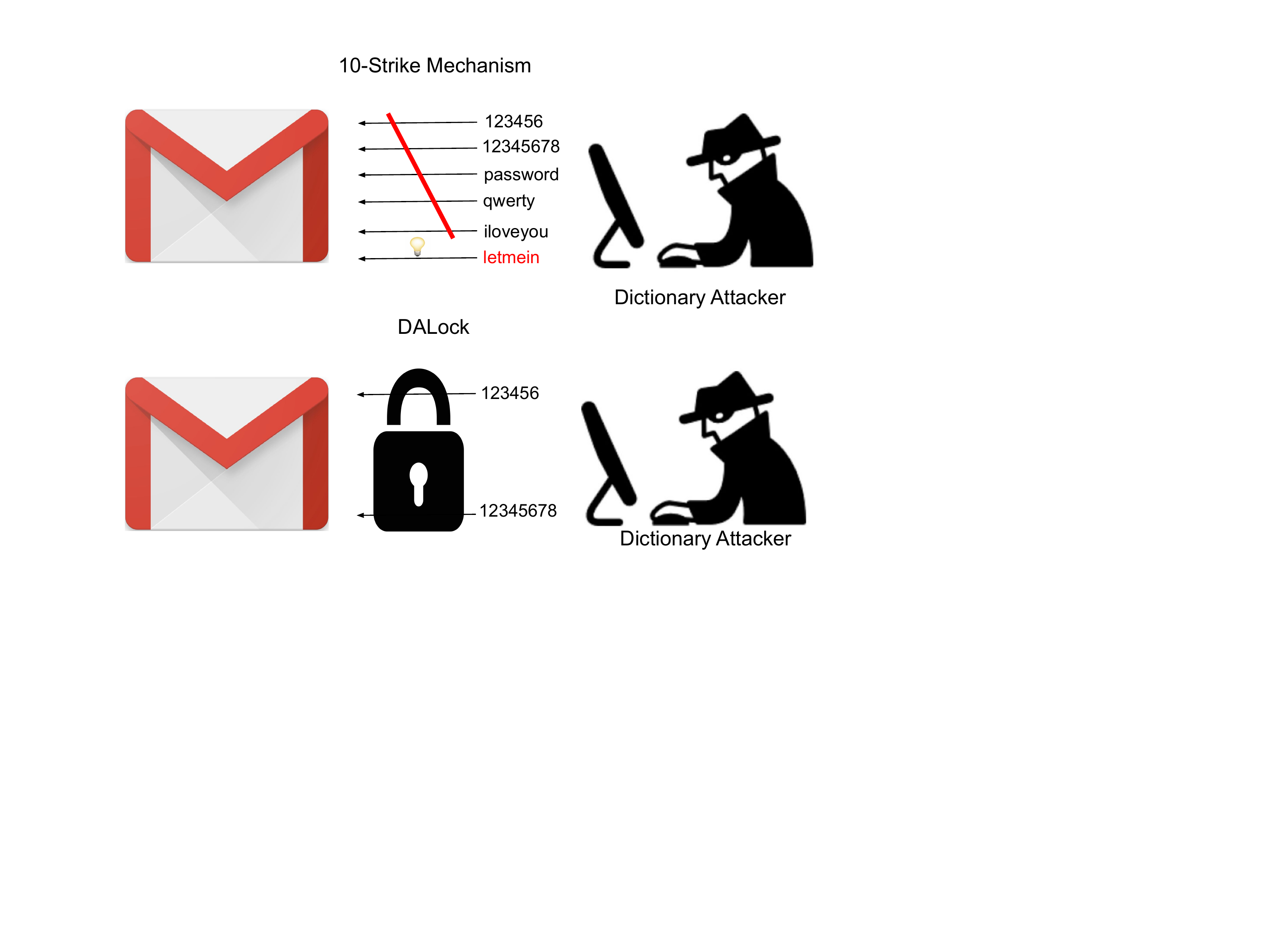}
		\caption{Security Comparison}\label{figure:introduction_security}
	\end{center}
	\vspace{-0.8cm}
\end{figure}

	\paragraph{Example 1: Usability} \textbf{Figure}~\ref{figure:introduction_usability} compares the usability of $\DALock$ with the standard $\strikeThreshold=3$ strikes mechanism. In this example scenario our user John Smith registers an account with the somewhat complicated password ``J.S.UsesStr0ngpwd!'' based on the story ``John Smith uses a strong password.'' Later when he tries to login he remembers the basic story, but not the exact password. Did he use his first name and his last name? With or without abbreviation? Did he add a punctuation mark at the end? Which letters are capitalized? If we use the 3-Strike mechanism John Smith will be locked out quickly e.g., after trying the incorrect password guesses ``JohnUseStrongPassword,'' ``JohnUsesStrongPassword'' and ``JohnUsesStrongpwd.'' However, since none of these passwords are overly popular $\DALock$ would allow our user to continue attempting to login until he recovers the correct password. 
	
	\paragraph{Example 2: Security} \textbf{Figure}~\ref{figure:introduction_security} compares $\DALock$ with the $K=10$ strike mechanism. In this scenario our user registers an account with a weak password ``letmein.'' Because the password is globally popular it is likely that an online attacker will attempt this password within the first $10$ guesses and break into the user's account. By contrast, $\DALock$ will quickly lockdown the account after the attacker submits two globally popular passwords. 
		
	We evaluate $\DALock$ empirically by simulating an authentication server in the presence of an online password attacker. We compare $\DALock$ with the the traditional $\strikeThreshold$ strikes mechanism for $\strikeThreshold \in \{3,10\}$. Our experiments show that when the hit count threshold $\hitCountThreshold$ is tuned appropriately that $\DALock$ significantly outperforms our $\strikeThreshold$-strike mechanisms. In particular, when user accounts are under attack we find that the fraction of accounts that are compromised is significantly lower for $\DALock$ than either $\strikeThreshold$ strikes mechanism --- even for the strict $\strikeThreshold=3$ strikes policy. We also evaluate the unwanted lockout rate of user accounts which are not under attack. We find that the unwanted lockout rate for $\DALock$ is much lower compare to $\strikeThreshold=3$-strikes mechanism. The unwanted lockout rate for $\DALock$ and the more lenient $\strikeThreshold=10$-strikes mechanism were comparable. A more detailed description of our experiments can be found in \textbf{section}~\ref{section:experimentalresult}.
		
	To deploy $\DALock$ we need a way to estimate the frequency of each incorrect login attempt to update $\hitCountThresholdOfU{u}$. We propose two methods for doing this: password strength meters (e.g. $\ZXCVBN$\cite{USENIX:Wheeler16}) and a differentially private count sketch data structure. Our empirical experiments show $\DALock$ outperforms traditional lockout mechisms with either approach. However, $\DALock$ performs best when we instantiate with a differentially private count sketch. On a positive note we show that even if the differentially private count sketch is trained on a small subset of user passwords that the estimates will still be high enough for $\DALock$ to be effective.

\section{Related Works and Backbrounds}\label{sec: relatedwork}

\subsection{Authentication Throttling} \label{related: Throttling}

\mypara{K-strike Mechanism} K-strike mechanism  is a straight-forrward implementation for authentication throttling. As its name suggests, throttling occurs when $\strikeThreshold$ consecutive incorrect login attempts are detected.  To reduce the cost of expensive overhead caused by unwanted throttling, Brostoff~\cite{brostoff2003ten} et.al suggest setting threshold $\strikeThreshold$ to be 10 instead of 3. They argue the increment risk is limited when strong password policy is enforced. However, this argument is challenged by empirical analyses of password composition policies~\cite{KSKMBCCE:SIGCHI11}\cite{BKPS:ACMEC13}. Many password composition policies do not rule out all low entropy password choices. For instance, it turns out that banning dictionary words does not increase entropy as expected.~\cite{KSKMBCCE:SIGCHI11} 

\mypara{Feature-Based Mechanism} To improve performance, modern throttling mechanisms\cite{sandhu2005system}\cite{gordon2014efficiently} often times use features such geographical location, IP-address, device information, and etc in addition to the correctness of attempting password. These features can be used to train sophisticated machine learning models to help distinguish between malicious and benign login attempts~\cite{NDSS:FJDBG16}. $\DALock$ takes an orthogonal approach and relies instead on the popularity of the password guesses. One can combine those models with a rigious throttling system for a better performance.

\mypara{Password-Distribution Aware Throttling} In an independent line of work Tian et al.~\cite{EuroSP:THS19} developed an IP-based throttling mechanism which exploits differences between the distribution of honest login attempts and attacker guesses. In particular, they propose to ``silently block'' login attempts from a particular IP address $a$ if the system detects too many popular passwords being submitted from that IP address. In more detail StopGuessing uses a data-structure called the binomial ladder filter~\cite{SchHer:MSR18} to (approximately) track the frequency $F(p)$ of each incorrect password guess. For each IP address the StopGuessing protocol maintains an associated counter $I_a = \displaystyle{\sum_{p \in \mathcal{P}} F(p)}$ where $\mathcal{P}$ is a list of incorrect password guesses that have been (recently) submitted from that IP address --- $I_a$ can be updated without storing $\mathcal{P}$ explicitly. Intuitively (and oversimplifying a bit) if $I_a > T$ then login attempts from address $a$ are silently blocked i.e., even if the attacker (or honest user) submits a correct password the system will respond that authentication fails. The authors also suggest protecting accounts with weak passwords by setting a user specific threshold $T(F(u_p))$ based on the strength $F(u_p)$ of the password $\PasswordOfU{u}$ of user $u$. Now if $I_a > T(F(u_p))$ then the system will silently reject any password from address $a$. Both StopGuessing and $\DALock$ exploit differences between the distribution of user passwords and attacker guesses. One of the key difference is that StopGuessing focuses on identifying malicious IP addresses (by maintaining a score $I_a$) while $\DALock$ focuses on protecting individual accounts by maintaining a ``hit-count'' parameter
 $\hitCountThreshold{u}$ for each user u. There are several other key differences between the two approaches. First, in $\DALock$ the goal of our frequency oracle (e.g., count-sketch, password strength meter) is to estimate the total fraction of users who have actually selected that particular password --- as opposed to estimating the frequency with which that password has been {\em recently} submitted as a incorrect guess. Second, $\DALock$ does not require silent blocking of login attempts which could create usability concerns if an honest user is silently blocked when they enter the correct password.  


\subsection{Passwords} \label{related: Passwords}  
\mypara{Password Distribution} Password distribution naturally represents the chance of success in the setting of statcial guessing attacks. Password distribution has been extensively studied since last decades\cite{FloHer:WWW07}\cite{DavKev:WWW12}. Using leaked password corpora\cite{Dataset:RockYou}\cite{Dataset:LinkedIn}\cite{SP:Bonneau12} is a straight forward way to describe the distribution of passwords. In recent works of  Wang et al.~\cite{EPRINT:WJHW14,TIFS17:WCWPXG,ESORICS:WanWan16} argue that password distributions follows Zipf's law i.e., leaked password corpora nicely fit Zipf's law distributions. Blocki et al.~\cite{SP:BloHarZho18} later found that Zipf's law nicely fits the Yahoo! password frequency corpus~\cite{SP:Bonneau12,NDSS:BloDatBon16}. \\
\mypara{Password Typos}  To test the usability of $\DALock$, it's crucial to reasonably simulate users' mistakes. Recent studies\cite{CCS:CWPCR17,SP:CAAJR16} from Chatterjee et. al have summarized probabilities of making (various of) typos when one enters his or her password based on users' studies. Based on the empirically measured data, they purposed two typo-tolerant authentication without sacrificing security. If fact, such mechanism has already been deployed in industry\cite{News:FacebookCaseSensitiveNews}\cite{News:AmazonTypo}. 

\subsection{Eliminating Dictionary Attacks} 

 \mypara{Increasing Cost of Authentcation} Pinkas and Sanders~\cite{CCS:PinSan02} proposed the use of puzzles (e.g., proofs of work or CAPTCHAs) as a way to throttle online password crackers. CAPTCHAs are hard AI challenges meant to distinguish people from bots~\cite{EC:vBHL03}. For example, reCAPTCHA~\cite{von2008recaptcha} has been widely deployed in online web services such as Google, Facebook, Twitter, CNN, and etc. Assuming that CAPTCHAs are only solvable by people, one can mitigate automated online dictionary attacks without freezing users' accounts~\cite{SP:BBFNJ10,CCS:BurMarMit11}. However, an attacker can always pay humans to solve these CAPTCHA challenges\cite{captchaSolver}. Increasingly sophisticated CAPTCHA solvers~\cite{NDSS:GYCZLT16,CCS:YTFZFX18} powered by neural networks make it increasingly difficult to design CAPTCHA puzzles that are also easy for a human to solve.  Golla~et al.~\cite{SOUPS:GBD17} proposed a fee-based password verification system where a small deposit is necessary to authenticate, which is refunded after successful authentication. A password cracker risks loosing its deposit if it is not able to guess the real password.

\mypara{Eliminating Popular Passwords} One mediation for dictionary attacks is eliminating the existence of weak or popular passwords. Schechter et. al~\cite{HTS:SchHerMit10} show that it is possible to forbid the existence of over popular password by maintaining the password distribution securely. Industrial solutions such as ``Have I been pwned?"~\cite{WebSite:HaveIBeenPwned} and ``Password CheckUp"~\cite{WebSite:GooglePasswordCheckUp} " prevent users to choose weak passwords based on data breaches. Password strength meters such as $\ZXCVBN$~\cite{USENIX:Wheeler16} are also been widely deployed to help users choosing stronger passwords.
 

\subsection{Privacy Perserving Aggregate Statistics Releasing.} \label{related: dp} $\DALock$ relies on the distribution of passwords to perform throttling. Storing/Releasing aggregate statistics naively often causes privacy leakage~\cite{arXiv:NarShm06}~\cite{UTA:NarShm08}. To answer the challenge, Cynthia Dwork purposed Differential Privacy~\cite{ECS:Dwork11} for aggregated data releasing. Informally speaking, differential private algorithm makes powerful adversaries unable of telling the existence of a record in the dataset. We defer the formal definition of differential privacy to section \ref{section:Prelinmaries-DiffernetialPrivacy}. Blocki et.al released the statics of Yahoo! dataset consist of 70 millions of passwords.~\cite{NDSS:BloDatBon16}. Recent work by Naor et at~\cite{CCS:NaoPinRon19} also demonstrates that releasing the distribution of password privately is feasible. In industry, Differential Privacy has been considered as the golden tool for various of tasks~\cite{AppleDP,AppleDPTeam,CCS:ErlPihKor14}.

\section{Preliminaries} \label{section:Prelinmaries}

\subsection{Count Sketch}\label{section:Prelinmaries-CountSketch} 
%
Count (Median) Sketch~\cite{ICALP:ChaCheFar02} and it's variants are widely in the tasks for finding frequent items such as popular passwords~\cite{CCS:NaoPinRon19}, homepage settings~\cite{CCS:ErlPihKor14}, and frequently used chat emojis~\cite{AppleDP}. In this work, we uses $\CountSketch$  as a tool for password popularity estimation. Formally speaking, we define Count Sketch as follows

\begin{definition}[Count Sketch~\cite{JoA:CorMut05}~\cite{ICALP:ChaCheFar02}]
A Count Sketch of state $\sigma: \mathsf{R}^{d\times w} \times \mathsf{R}$ is represented by a two-dimensional $d\times w$ array counts $\CountSketch$, a total frequency counter $\CountSketchCounter$, and d + 1 hash functions ($h_1, \ldots, h_d, h_{\pm}$) chosen uniformly at random from a pairwise-independent family. 
\begin{center}
$h_1 \cdots h_d$ : $\{pw\} \rightarrow \{1\cdots w\}$\\
$h_{\pm} : \{pw\} \rightarrow \{1, -1\}$\\
\end{center}
\end{definition}

In this work we consider the following four classic APIs for Count Sketch: initialize, Add, Estimate, and TotalFreq. Additionally, we consider an extra operation DP which is used to construct differentially private Count Sketch from a standard one.

\mypara{$\sigma_{0} \leftarrow$ Initialize($d,w$)}: This API initialize and return a Count Sketch of state $0^{d\times w}\times 0$, i.e. an all zero table.

\mypara{$\sigma_{new} \leftarrow $ Add($pw, \sigma$):} Add operation updates the stored frequency count password $pw$ based on a $\CountSketch$ state $\sigma$, and outputs the updated state $\sigma_{new}$.

In addiition, given a multiset $\SampledData{\AllUser} = \{pw_1,...,pwd_N\}$, we use the following notation $\sigma_{\SampledData{\AllUser}} =  Add(\SampledData{\AllUser},\sigma) \\= Add(pw_1, Add(\{pw_2,...,pw_N\},\sigma)$ to ease presentation. Further more, we omit subscript $\AllUser$ and simply use $\sigma$ to denote $\sigma_{\SampledData{\AllUser}}$.

\mypara{Estimate($pw,\sigma$)}: This interface returns the estimated frequency of password $pw$ based on the given Count Sketch State $\sigma$.

To implement $\DALock$ with high accuracy, we want the estimator has the following correctness Property: $\EstimateF{pw}{\sigma} \approx \TrueFInD{pw}{\SampledData{\AllUser} }$. 

\mypara{TotalFreq($\sigma$)}: This opeartion returns the total number of passwords based on state $\sigma$.

Based on the above definition, we denote the \emph{estimated popularity of password} $pw$ by $\sigma$ with $\EstimateP{pw}{\sigma} = \frac{\EstimateF{pw}{\sigma}}{\text{TotalFreq}(\sigma)}$. For the rest of discussions, we sometimes omit $\sigma$ when there is no ambiguity to simplify presentation. e.g. $\EstP{pw}$ = $\EstimateP{pw}{\sigma}$. In addition, we allow the above APIs to take a set of passwords as argument and return the summed results. i.e.. $\EstP{S} = \displaystyle{\sum_{pw \in S} \EstP{S}}$. 

\mypara{$\sigma_{dp}$ $\leftarrow$  DP($\epsilon, \sigma$)}:This function outputs an differentially private state $\sigma_{dp}$ of $\sigma$ with privacy budget $\epsilon$.

\subsection{Differential Privacy} \label{section:Prelinmaries-DiffernetialPrivacy} 
Differential Privacy ~\cite{ECS:Dwork11} is one of the industrial golden standard tools for private aggregated statical releasing. Intuitively speaking, individual record has limited impact on computing the final results to be published if one applies differential. For instance, consider the following two password datasets: A password dataset$\SampledData{\AllUser}$ consists of users' passwords , and it's neighboring dataset $\SampledData{\AllUser - pw_u}$ obtained by removing user u's password $pw_u$ from $\SampledData{\AllUser}$. Differential privacy guarantees that \emph{with high probability} the published results based on $\SampledData{\AllUser}$ and $\SampledData{\AllUser - pw_{u}}$ are the same. Therefore, adversary is not able to infer the existence of $u$ in dataset $\SampledData{\AllUser}$.

In this work, we adopt differential privacy Count Sketch to reduce the risk of privacy leakage. Based on the our notion of Count Sketch, one can define differential privacy as follows
\begin{definition}[{$\epsilon$-Differential Privacy~\cite{ECS:Dwork11}}] \label{def:diff}
A randomized mechanism $\mathcal{M}$ gives $\epsilon$-differential privacy if for any pair of neighboring datasets $\SampledData{\AllUser}$ and $\SampledData{\AllUser}'$, and any $\sigma \in \mathit{Range}(\mathcal{M})$,
$$\Pr{\mathcal{M}(\SampledData{\AllUser})=\sigma} \leq e^{\epsilon}\cdot \Pr{\mathcal{M}(\SampledData{\AllUser}')=\sigma}.$$
\end{definition}

 We consider two datasets $\SampledData{\AllUser}$ and $D'$ to be neighbors i.f.f. either $\SampledData{\AllUser}=\SampledData{\AllUser}' + pw_u$ or $\SampledData{\AllUser}'=D + pw_u$, where $\SampledData{\AllUser} +pw_u$ denotes the dataset resulted from adding the tuple $pw_u$(a new password) to the dataset $\SampledData{\AllUser}$. We use $\SampledData{\AllUser}\simeq \SampledData{\AllUser}'$ to denote two neighboring datasets. This protects the privacy of any single tuple, because adding or removing any single tuple results in $e^{\epsilon}$-multiplicative-bounded changes in the probability distribution of the output. If any adversary can make certain inference about a tuple based on the output, then the same inference is also likely to occur even if the tuple does not appear in the dataset.

\mypara{Laplace Mechanism.} 
The Laplace mechanism is a classic tool achieve differential privacy. It computes a differential private state $\sigma$ on the dataset $\SampledData{\AllUser}$ by adding a random noise. The magnitude of the noise depends on $\mathsf{GS}_\sigma$, the \emph{global sensitivity} or the $L_1$ sensitivity of $\sigma$.  $\mathsf{GS}_\sigma$ quantify the maximum impact on $\sigma$ by adding or removing any record.

\mypara{Differentially Private Count Sketch} Given a $\CountSketch$ of state $\sigma$, adding (removing) any password $pw_u$ to(from) it can result in at most d + 1 changes for $l_1$ norm. Because each $pw_u$ contributes to d entries in the $d \times w$ table and total count. Therefore, To release $\sigma$ with privacy budget $\epsilon$, it suffices to add $\Lap(\frac{d+1}{\epsilon})$ to all entries in $\sigma$. 

\mypara{Privacy-Preserving Password Corpus Relasing Mechanism} Naor et.al\cite{CCS:NaoPinRon19}  purposed an algorithm to release password distribution using local differential privacy. In our work, we focused on a centralized version of differential privacy which is expected to have less noisy compare to local setting. StopGuessing\cite{EuroSP:THS19} uses a binomial ladder to identify ``heavy hitters'' (popular passwords), though the data-structure does not provide any formal privacy guarantees such as differential privacy. The data-structure is not suitable for $\DALock$ as it provides a binary classification i.e., either the password is a ``heavy hitter'' or it is not. For $\DALock$ requires a more fine grained estimate of a passwords popularity. 


\subsection{Notation Summary}
In this section, we summarize frequently used notations in this paper across all sections in \textbf{Table} ~\ref{table: notation}.  For a password $pw \in \AllPassword$ we use $\TrueP{pw}$ to denote the probability each user selects the password $pw$. We assume that there is some underyling distribution over user passwords and use $\TrueP{pw}$ to denote the probability of the password $pw \in \AllPassword$. It will be convenient to assume that all passwords $ \AllPassword = \{pw_1,pw_2,\ldots \}$ are sorted in descending order of probability i.e., so that $\TrueP{pw_1} \geq \TrueP{pw_2} \ldots $. We will use $pw_r = \TrueP{pw_r}$ to denote the probability of the $r$th most likely password in the distribution. 

 We use  $\AllUser = \{u_1,\ldots,u_N\}$ to denote a set of  $N$ users and $D_\mathcal{\AllUser} \subseteq \AllPassword$ is a multiset of user passwords i.e., $D_\mathcal{\AllUser} = \{pw_{u_1},\ldots,pw_{u_N}\}$. We typically view $D_\mathcal{\AllUser}$ as $N$ independent samples from an underlying distribution over $\AllPassword$ and write $\TrueF{pw, \mathcal{D}_U} = \left| \left\{ i ~:~pw_{u_i} = pw \right\} \right|$ to denote the number of times the password $pw$ was observed in our sample. We often omit $D_\mathcal{\AllUser}$ when the dataset is clear from context and simply write $\TrueF{pw}$. 
 
 We remark that $\TrueP{pw} = \frac{\mathbb{E}\left[ \TrueFInD{pw}{ \mathcal{D}_U}\right]}{N}$ and thus for popular passwords we expect that the estimate $\TrueP{pw} \approx  \frac{\TrueF{pw, \mathcal{D}_U}}{N}$ will be accurate for sufficiently large $N$. However, because the underlying password distribution is unknown and an authentication server cannot store a plaintext encoding of $D_\mathcal{\AllUser}$ we will often use other techniques to estimate  $\TrueP{pw}$ and/or $\TrueF{pw, \mathcal{D}_U}$. In particular, we consider a Count (Median) Sketch data structure $\CountSketch$ trained on $\mathcal{D}_U$ (or a small subsample of $\mathcal{D}_U$) which allows us to generate an estimate $\EstP{pw}$ for the popularity of each password. Similarly, we can also use password strength meters to estimate  $\TrueP{pw}$.

\begin{table}[htb]
	\begin{tabular}{|l|l|l|}\hline
		Notation      & Description                                                                   \\\hline
		$\Adversary$  & \underline{$\Adversary$}dversary                            \\\hline
		$\AllUser$ & The set of 	{$\AllUser$}sers           \\\hline
		$u$           & A user  $u \in \AllUser$                                                    \\\hline
		$\AllPassword$ & The set of all potential user \underline{$\AllPassword$}asswords \\\hline
		$\SampledData{\AllUser} \subseteq \AllPassword$ & a multiset of $N$ sampled passwords  for\\
																   		& users $u_1,\ldots,u_N \in \AllUser$ \\ \hline
		$\PasswordOfU{u}$         & User $u$'s password   \\\hline
		$\RankRPassword{r}$         & The $r$'th most likely password in $\SampledData{\AllUser} \subseteq \AllPassword$ \\\hline
		$\CountSketch$ & \underline{C}ount (Median) \underline{S}ketch data structure\\    \hline    	
		$\TrueFInD{pw}{\SampledData{\AllUser}}$ & \underline{F}requency of password $pw$ in dataset $\SampledData{\AllUser}$ \\\hline
		$\TrueP{pw}$ & Empirical probability of password $pw$  \\\hline        
		$\EstF{pw}$ & Estimated frequency of password $pw$\\\hline
		$\EstP{pw}$ & Estimated probability of password $pw$  \\\hline                    
		$\hitCountThreshold$ & Hit count threshold  \\\hline 
		$\hitCountThresholdOfU{u}$ & Remainining hit count threshold of user u. \\&The account gets locked out if $\hitCountThresholdOfU{u}$ reaches $\hitCountThreshold$\\\hline
		$\strikeThreshold$ & traditional $K$-strike threshold. \\\hline
		$\strikeThresholdOfU{u}$ & Remaining strike threshold on $u$'s account. \\&The account gets locked if $\strikeThresholdOfU{\AllUser}$ exceeds $\strikeThreshold$. \\\hline
	\end{tabular}
	\caption{Notation Summary}\label{table: notation}
	\vspace{-1cm}
	
\end{table}



\section{The $\DALock$ Mechanism}\label{sec:DALockAlgorithm} 
In this section, we present the $\DALock$ mechanism, discuss how $\DALock$ might be implemented and the strategies that an attacker might use when $\DALock$ is deployed. Intuitively, $\DALock$ punishes incorrect password guesses more harshly if the guessed password $pw$ is overly popular since an attacker will want to submit popular password guesses to maximize his chances of cracking the user password.

\subsection{$\DALock$} 
In the classic $K$-strike throttling mechanism we keep track of a parameter $\KOfU$ which tracks the number of consecutive incorrect login attempts for each user $u$. After each consecutive login attempt the parameter is $\KOfU$ and the parameter $\KOfU$ is reset to $0$ whenever $u$ authenticates successfully. If we ever have $\KOfU \geq \strikeThreshold$ then throttling mechanism kicks in and the authentication server will lock down the account until the user takes some action\footnote{For example, the user might be asked to  resetting his password via e-mail or wait for some fixed amount of time. In some settings the user might simply be asked to solve a CAPTCHA challenge. The latter approach has some usability advantages and security drawbacks e.g., a malicious password might pay to solve the CAPTCHA challenges so that he can continue attempting to guess the user's password}.

The key-idea behind $\DALock$ is to additionally  maintain an extra ``hit count'' variable $\hitCountThresholdOfU{u}$ for each user $u$. Intuitively, $\hitCountThresholdOfU{u}$ measures the total probability mass of all incorrect password guesses submitted for user $u$. Initially, when a new user registers we will have $\hitCountThresholdOfU{u}=0$ (and $\KOfU =0$). After each attempted login with an incorrect  password $pw \neq pw_u$ the hit count is incremented so that  $\hitCountThresholdOfU{u}+= \EstP(pw)$. Here, $\EstP(pw)$ denotes an estimate for the probability of the password $pw$ so that incorrect passwords are punished more severely when $pw$ is an overly popular password. Unlike the consecutive strikes parameter $\KOfU$ which is reset to $0$ after each successful login, the hit count parameter can only be incremented. $\DALock$ throttles $u$'s account if the ``hit count'' exceeds $\hitCountThreshold$ (i.e., $\hitCountThresholdOfU{u} \geq \hitCountThreshold$) or if there are too many consecutive mistakes (i.e., $\strikeThresholdOfU{u} \geq \strikeThreshold$)  For example, suppose that the (estimated) probability of the passwords ``aaa," ``bbb" and ``ccc'' were 3\%, 1.7\% and 0.8\%. If a user registers with a password ``ddd'' and then attempts to login with the previous three passwords then $\hitCountThresholdOfU{u}$ will be set to $0.055=0.03+0.017+0.008$. 

Each time the user (or attacker) attempts to login with a password $pw$ the response will either be (1) ``locked'' if $\hitCountThresholdOfU{u} \geq \hitCountThreshold$ or if $\strikeThresholdOfU{u} \geq \strikeThreshold$, (2) ``correct'' if the the guessed password matches the user password i.e., $pw = pw_u$\footnote{To ease presentation we omit the description of the password hashing algorithm when we describe the authentication server. In practice, we recommend that the authentication server only stores salted password hashes using a moderately expensive key derivation function to increase guessing costs for an offline attacker.} or (3) ``incorrect password'' otherwise. We demonstrate the login flow in \textbf{Algorithm}~\ref{algorithm:DALock}, Appendix. We remark that the authentication server could intentionally blur this distinction between cases (1) and (3), but that this comes at a usability cost e.g., an honest user would be annoyed if they were repeatedly informed that their password is incorrect whenever the account is actually locked.

{\noindent \bf Remark:} One could optionally consider initializing the hit count parameter $\hitCountThresholdOfU{u}$ based on the strength of the user's password. For example, if $u$ registers with a weak password then we might initialize $\hitCountThresholdOfU{u} = \hitCountThreshold/2$ for stronger protection i.e., so that the account is locked down faster. Similarly, a user with a strong password might be awarded by setting $\hitCountThresholdOfU{u} = \hitCountThreshold$. However, because $\hitCountThresholdOfU{u}$ and $\strikeThresholdOfU{u}$ are stored on the authentication server this would leak information about the strength of $pw_u$ to an offline attacker e.g., if an offline attacker sees that $\hitCountThresholdOfU{u} = \hitCountThreshold/2$ he might reasonably infer that the user picked a weak password. \footnote{One could potentially avoid storing $\hitCountThresholdOfU{u}$ unencrypted if one is willing to implement a silent lockout policy where the user cannot distinguish between an incorrect guess and a locked account, but we wish to avoid solutions that blur this distinction.}

\subsection{$\DALock$ Authentication Server} 
To implement $\DALock$ we need an efficient way to estimate the probability $\EstP(pw)$ of each incorrect password $pw$. We consider several instantiations of this frequency oracle. One option is to use password strength meters such as $\ZXCVBN$ or more sophisticated password cracking models e.g., Markov Models, Probabilistic Context Free Grammars, or Neural Networks. Another naive approach would be to simply maintain a plaintext list of all user passwords along with their frequencies. However, this approach is inadvisable due to the risk of leaking this plaintext list. Herley and Schechter~\cite{HTS:SchHerMit10} proposed the use of the Count-Sketch data-structure which would allow us to estimate the frequency of each password without explicitly storing a plaintext list although there are no formal privacy guarantees to this approach. We chose to adopt a Differentially Private Count-Median-Sketch. The authentication server initializes the Count-Sketch $\sigma_{dp}$ $\leftarrow$  DP($\epsilon, \sigma$) by adding Laplace Noise to preserve $\epsilon$-differential privacy and each time a new user $u$ registers a new password  $pw_u$ would be added to the Count Sketch.


We remark that maintaining a Differentially Private Count-Sketch has many other potentially beneficial applications e.g., one could use the Count-Sketch to ban weak passwords~\cite{HTS:SchHerMit10} and/or to help identify IP addresses associated with malicious online attacks~\cite{EuroSP:THS19}. One disadvantage is that the attacker will also be able to view the Count-Sketch data-structure if the data-structure is leaked. The usage of differential privacy helps to minimize these risks. Intuitively, differential privacy hides the influence of any individual password ensuring that an attacker will not be able to use the Count-Sketch data-structure to help identify any unique passwords. However, an attacker may still be able to use the data-structure to learn that a particular password is globally popular (without linking that password to a particular user). We argue that this is not a major risk as most attackers will already know about globally popular passwords e.g., from prior breaches.

\section{Experimental Design} 
We evaluate the performance of $\DALock$ through an extensive battery of empirical simulations. In this section we describe the modeling choices we made when designing our experiments. To simulate the authentication ecoystem we need to simulate the behavior of honest users,  the authentication server running $\DALock$  and an online attacker. 

 Briefly, when simulating users we need to model the distribution over user passwords, the distribution over honest login mistakes (e.g., typos or recall errors) as well as the user's login schedule. When simulating the distribution over user passwords we use three empirical datasets (RockYou, Yahoo!, and LinkedIn) to define the underlying password distribution. We use a Poisson arrival process to model the frequency of user login attempts~\cite{AC:BloBluDat13}. Our model for user mistakes is informed by recent empirical studies of password typos~\cite{SP:CAAJR16,CCS:CWPCR17} and is augmented to simulate other mistakes i.e., recall errors.  The key question to answer for simulating an authentication server running $\DALock$ is how the (password) frequency oracle $\EstP{\cdot}$ is implemented. We consider two concrete implementations: password strength models (e.g., $\ZX$, Markov Models, Neural Networks) and (differentially private) count sketch. When simulating the attacker we consider an untargetted one who knows the distribution over user passwords as well as the DALock mechanism --- including the frequency oracle  $\EstP{\cdot}$ . We leave the question of tuning DALock to protect against targetted online attackers~\cite{CCS:WZWYH16} as an important direction for future research. We elaborate on each of these key model components below.  We begin by with an overview of the empirical datasets $\SampledData{\AllUser}$ that we used in our experiments.

\subsection{Experimental Datasets}\label{section:experiment:experiment_dataset} 
	In this work we use three publicly available password datasets: LinkedIn\cite{Dataset:LinkedIn}, {RockYou\cite{Dataset:RockYou}, and Yahoo\cite{SP:Bonneau12,NDSS:BloDatBon16}. In addition, we construct datasets via subsampling and password banning mentioned in \textbf{section}~\ref{section:ExperimentDesign-subsection:SimulateServer}. We summarize the characteristics of datasets in Table \ref{Table:dataset} to help reader catch the essence. 

	\textbf{RockYou}\cite{Dataset:RockYou} is a plaintext password corpus contains 14,341,564 unique passwords from a data leakage in 2009. RockYou stored user data in unencrypted database and leaked the passwords due to SQL-injection attacks. 

	\textbf{LinkedIn}\cite{Dataset:LinkedIn} is a plaintext password corpus (partially) recovered constructed from a leakage in 2012. While the passwords were originally hashed LinkedIn was using a weak (unsalted) password hashing algorithm and almost all of the passwords in the dataset have been cracked. The dataset we used has approximately $68$ million passwords. We remark that the actual size of the leak is larger and that there is a larger (differentially private) frequency corpus based on $174+$ million LinkedIn passwords~\cite{harsha2020bicycle} that is publicly available. However, this dataset does not include any plaintext passwords. We chose to use the smaller dataset in our experiments so that we could evaluate with frequency oracles based on password models (e.g., $\ZXCVBN$, PCFGs, Neural Networks). 
	

	\textbf{Yahoo!} The Yahoo! frequency corpus is a sanitized password frequency dataset collected~\cite{SP:Bonneau12}   with permission from Yahoo! It consists of anonymized password histograms representing almost 70 million Yahoo! users who logged into their account during a $48$ hour window in May, 2011. Yahoo! later authorized the public release of a differentially private version of this dataset~\cite{NDSS:BloDatBon16}. We remark that this frequency corpus does not contain any plaintext passwords so we did not use password strength models in our experiments involving the Yahoo! dataset.

\begin{table}[h]
\begin{tabular}{|c|c|c|c|}
\hline
Dataset     & Unique Passwords & Accounts & $\TrueP{pw_1}$\\ \hline
LinkedIn &  6,840,885              &68,361,064                    &1.53\%                         \\ \hline
Yahoo    & 33,895,873                     &  69,301,337            & 1.1\%                   \\ \hline
RockYou  & 14,341,564                & 32,603,388                      & 0.89\%                    \\ \hline
\end{tabular}
\caption{Summary of dataset}\label{table: datasetsummary}
\label{Table:dataset}
	\vspace{-0.90cm}
\end{table}
\noindent \textbf{Remark:} In this work, we assume each dataset (approximately) reflects the actual distribution due to their tremendous size. To ease presentation, we refer ``actual distribution" to the datasets described in this section. We do acknowledge there could be misalignment between the above ones and their undelying true distributions. But we focus on analyzing popular passwords where empirical distribution and the real distribution will be very similar.

\mypara{Ethics} Some of the datasets we used (LinkedIn\cite{Dataset:LinkedIn} and  {RockYou\cite{Dataset:RockYou}) contain passwords that were previously stolen and subsequently leaked online. The use of this data raises important ethical considerations. We remark that the password lists are already publicly available online so our use of the data does not exacerbate the prior harm to users. We did not crack any new user passwords. Furthermore, the data we use has been cleaned of all identifying information beyond the passwords themselves.  In summary, we believe that our use of the leaked data will not exacerbate prior harm to users, and that the lockout mechanism we develop and evaluate may help to protect user passwords in the future.

\subsection{Modeling Users} \label{section:ExperimentDesign-subsection:SimulateUser}
Our model to simulate the behavior of honest users consists of three key components: user password selection, login frequency, and mistake model. 

\subsubsection{Simulating Users' Choice of Password}\label{section:ExperimentDesign-subsection:SimulateUser-subsubsection:SimulatePasswordChoice}
In each simulation we fix a dataset which is used to simulate user password selection. We use three large empirical password datasets: RockYou, LinkedIn and Yahoo! In particular, a dataset consists of a multiset  $\SampledData{\AllUser} = {pw_1,...,pw_N}$ of $N$ user passwords which can be compressed into pairs $(pw,  \TrueFInD{pw}{\SampledData{\AllUser} })$ --- recall that $\TrueFInD{pw}{\SampledData{\AllUser} }$ denotes the number of time the password $pw$ occurs in the dataset $\SampledData{\AllUser}$.   See  \textbf{Section}~\ref{section:ExperimentDesign-subsection:SimulateUser-subsubsection:SimulatePasswordChoice} for a more details about each dataset. Each of these datasets $\SampledData{\AllUser} $ induces an empirical distribution over user passwords where the probability of sampling each password $pw$ is simply $\TrueFInD{pw}{\SampledData{\AllUser}}/N$\footnote{In our analysis we will assume that the empirical distribution is the {\em real distribution} over user passwords and which is also known to the attacker. Given a password datƒaset $\SampledData{\AllUser} $  sampled from the real distribution over user passwords we remark that when analyzing online attacks we focus on popular passwords where empirical distribution and the real distribution will be very similar. }.  Each simulated user $u$ in our experiment samples $6$ passwords from this empirical distribution and registers with the first password. Intuitively, the five extra sampled passwords will  be used to help simulate recall errors e.g., they represent the user's passwords for other websites. 

We remark that the Yahoo! dataset~\cite{SP:Bonneau12,NDSS:BloDatBon16} only contains frequencies without actual passwords i.e., instead of recording the pair $(pw,  \TrueFInD{pw}{\SampledData{\AllUser} })$ the dataset simply records $\TrueFInD{pw}{\SampledData{\AllUser} }$ . We generate a complete password dataset by designating a unique string for each password. As we avoid using password models like $\ZX$ to analyze $\DALock$ with the Yahoo! dataset since frequency estimation requires accesss to the original passwords. However, we are still able to evaluate $\DALock$ with the Yahoo! dataset using the Count-Sketch frequency oracle. 

\mypara{Banlists} We additionally consider the setting where the authentication server chooses to ban users from selecting the top $B$ passwords e.g., $B=10^4$ passwords. We use the normalized probabilities model~\cite{BKPS:ACMEC13} to simulate user password selection under such restrictions. In this model we simply use rejection sampling to avoid sampling one of the top $B$ passwords. Equivalently, we can let $\SampledData{\AllUser, B}$ denote the dataset $\SampledData{\AllUser}$ with the $B$ most common passwords removed and sample from the empirical distribution corresponding to the updated dataset $\SampledData{\AllUser, B}$.

\subsubsection{Simulating user's login patterns}\label{section:ExperimentDesign-subsection:SimulateUser-subsubsection:SimulateLoginPattern} 
To simulate each user we need to model the frequency with which our honest user attempts to login to the authentication server. In particular, we aim to simulate the login behavior over a 180 day time span. For each user $u$ we want to generate a sequence $0 < t_1^u < t_2^u < \cdots <  4320 = 180\times24$  where each $t_i^u \in \mathbb{N}$ represents the time (hour) of the $i$th user visit. Following prior work (e.g., see \cite{AC:BloBluDat13,CCS:KogManBon17}) we use a Poisson arrival process to generate these login times. The Poissuon arrival process is parameterized by an arrival rate $T_u$ (hours) which encodes the expected time between consecutive login attempts $T_u = \mathbb{E}[t_{i+1}-t_i]$, and the arrival process is memoryless so the actual gap $t_{i+1}-t_i$  is independent of $t_i$. Since some users are more active than others we pick a different arrival rate $T_u$ for each user $u$ where each $T_u$ is sampled uniformly random from $\{ 12, 24, 24 \times 3, 24 \times 7, 24 \times 14, 24 \times 30\}$. The parameter $T_u = 12$ (hours) corresponds to users who visit multiple times per day on average, while the parameter $T_u = 24 \times 30$ corresponds to a user who visits the site once per month. For each user $u$ we use the Poisson arrival process with parameter $T_u$ to generate the sequence of visits $0 < t_1^u < t_2^u < \cdots <  4320 = 180\times24$   over a time span of $180$ days ($4320$ hours). Each time a user visits we assume that they will continue attempting to login until they succeed or get locked out. 

We remark that we do not simulate a client device which automatically attempts to login on the user's behalf. It may be desirable to have the authentication server store the (salted) hash of the user's previous password(s) so that we can avoid locking the user's account in settings where a client device might repeatedly attempt to login with an outdated password. Alternatively, the authentication server could store an encrypted cache of incorrect login attempts using public key cryptography where each incorrect login attempt $pw_u' \neq pw_u$ would be encrypted with a public key $pk_u$ and stored on the authentication server. The encrypted cache could only be decrypted when the user authenticates with the correct password\footnote{Unlike the public encryption key $pk_u$, which would be stored on the authentication server, the secret key $sk_u$ would only be stored in encrypted form i.e., the server would store $c_u = \mathbf{Enc}_{K_u}(sk_u)$ where $K_u = \mathbf{KDF}(pw_u)$ is a symmetric encryption key derived from the user's password. }. The encrypted cache could be used as part of a personalized typo corrector~\cite{CCS:CWPCR17} and could also be used to avoid penalizing repeat mistakes~\cite{CCS:CWPCR17,EuroSP:THS19}. One potential downside to this approach is that the cache might inadvertantly contain credentials from other user accounts making cached data valuable to the attacker. More empirical study would be  needed to determine the risks and benefits of maintaining such a cache.




\subsubsection{Simulating User Mistakes}\label{section:ExperimentDesign-subsection:SimulateUser-subsubsection:SimulateUserMistake} 
The last component of our user model is a mechanism to simulate user mistakes during the authentication process. Our model relies upon recent empirical studies of password typos~\cite{CCS:CWPCR17,SP:CAAJR16} and additionally incorporates other common user mistakes e.g., recall errors. The aforementioned studies show that roughly $7.5\%$ of login attempts are mistakes and at least $68\%$ of these mistakes are (most likely) typos i.e., within edit distance $2$ of the original passwords.  
Accordingly, we set the mistake rate to be $7.5\%$ for simulation. When simulating each login attempt the user will enter the correct password with probability $92.5\%$. Otherwise, if the user makes a mistake we simulate a typo with probability $68\%$ and we simulate a recall error with probably $32\%$. To simulate a recall error we randomly select one of the user's five alternate passwords to model a user who forgot which of his passwords was associated with this particular account --- if the user recalls the wrong password they might additionally miss-type it (with probability $0.075\cdot 0.68$).  We refer an interested reader to the appendix for a more detailed discussion of our mistake model including a flow chart (see Figure~\ref{figure:flowChartTypo}) and more finegrained typo statistics (see  Table~\ref{Table:TypoTypes}).

We remark that we do not attempt to simulate a user who completely forgets his password. Of course we expect that this will occasionally happen in reality. However, we observe that a user who forgets his password will {\em always} need to reset it regardless of the throttling mechanism adopted by the authentication server. 

\subsection{Modeling the Authentication Server}\label{section:ExperimentDesign-subsection:SimulateServer} 
We model an authentication server running $\DALock$ with various parameters $\strikeThreshold$ and $\hitCountThreshold$ for the strike count and hit count. Each time a user $u$ (or attacker pretending to be $u$) attempts to login the authentication server updates the parameters $\hitCountThresholdOfU{u}$ and $\strikeThresholdOfU{u}$ accordingly following the $\DALock$ mechanism. We remark that when $ \hitCountThreshold = \infty$ that the authentication server is running the classical $ \strikeThreshold$-strikes lockout policy. To deploy $\DALock$ with a finite hitcount parameter $ \hitCountThreshold$ an authentication server needs to use a frequency oracle to update the hit count after each incorrect login attempt.  In this work we consider two concrete approaches the authentication server might adopt: (differentially private) Count Sketch estimator and Password Strength Models. We use $\EstimateP{pw}{\Estimator}$ to denote the estimated popularity (probability) of a password $pw$ using the estimator $\Estimator$ e.g., given a Count-Sketch $\sigma$ we would use  $\EstimateP{pw}{\sigma} = \frac{\mathbf{Estimate}(pw,\sigma)}{\mathbf{TotalFreq(\sigma)}}$. We remark that the authentication server might (optionally) chose to ban overly popular passwords to flatten the password distribution to protect user accounts against online attackers  \cite{HTS:SchHerMit10}. If the authentication server adopts such a policy then the frequency oracle would need to be adjusted accordingly to model the new password distribution.

\subsubsection{Differentially Private Count Sketch Estimator} 
The first instantiation of  $\EstimateP{\cdot}{\cdot} $ we consider is to build a Count Sketch Estimator $\sigma_{\SampledData{\AllUser}} = \Add{\SampledData{\AllUser}}{\sigma} $ from our dataset $\SampledData{\AllUser} $ of user passwords. To build a Count Sketch in practice the authentication server would update the Count Sketch with the new password each time a user registers \footnote{The Count Sketch instantiations we consider would also support a Remove operation which would allow the authentication server to handle password updates efficiently}. There are several issues to consider when deploying the Count Sketch estimator: memory efficiency, privacy, sample size and accuracy. 

\textbf{Memory Efficiency} We instantate the Count Sketch with parameters $d=5$ and $w=10^6$ so that the entire datastructure requires just $20$ MB of space which easily fits in RAM. 

\textbf{Privacy} As we discussed earlier one concern about storing a Count Sketch $\sigma_{\SampledData{\AllUser}} $ on the authentication server is that an offline attacker might steal this file and use the data-structure to help identify user passwords. For example, if our user John Smith selects (resp. does not select) the password ``J.S.UsesStr0ngpwd!'' then we would expect that the true frequency of this password is $\TrueFInD{pw}{\SampledData{\AllUser} }=1$ (resp. $\TrueFInD{pw}{\SampledData{\AllUser} }=0$). If the Count Sketch estimator is overly accurate then the attacker would be able to learn that one user (most likely John Smith) picked this password. Without a way to address these privacy concerns an organization might be understadibly wary to deploy a Count Sketch estimator.

To address these  privacy concerns we consider an $\epsilon$- differentially private estimator $\sigma_{dp}$ = \textbf{DP($\epsilon,\sigma$)} in our experiments. During initialization we add laplace noise to each of the cells in the Count Sketch where the noise parameter scales with $d/\epsilon$. In our above example, differential privacy ensures that --- up to a multiplicative advantage $e^{\epsilon}$ --- an attacker cannot use the count sketch to distinguish between a dataset in which John Smith did (resp. did not) pick the password ``J.S.UsesStr0ngpwd!' We remark that lower values of $\epsilon$ correspond to stronger privacy guarantees e.g., we use $\epsilon=\infty$ to denote the case with no differential privacy guarantees. In most of our experiments we use a small privacy parameter $\epsilon=0.1$ which is much smaller than the privacy parameters used in most prior deployments of differential privacy e.g.,  \cite{NDSS:BloDatBon16,AppleDPTeam,CCS:ErlPihKor14}. 

\textbf{Sample Size and Accuracy} In general the accuracy of a Count Sketch increases with the size of the password dataset. Suppose that the organization does not have millions of users or the that the sample size is decreased because the organization allows users to ``opt-in'' to the (differentially private) count sketch. One natural question is whether a smaller organization would be able to deploy a Count Sketch  to obtain reliable frequency estimates. We investigate this question by subsampling smaller datasets to train the Count Sketch. Given a set $\AllUser$ of $N$ users we use $\AllUser_{r\%}$ to denote a randomly subsampled set of $r\%$ of users. We use $\SampledData{\AllUser_{r\%}}$ to denote the corresponding subsampled password dataset $\sigma_{r\%} = \Add{\SampledData{\AllUser}}{\sigma} $ to denote the Count Sketch trained on the subsampled data. The question is whether $\sigma_{r\%}$ can be as effective as $\sigma$ for deploying $\DALock$. 

In our experiments we consider the following sampling rates: 1\%, 5\%, and 10\%. We find that even when $r=1\%$ the Count Sketch $\CountSketch$ trained on $\SampledData{\AllUser_{1\%}}$ is sufficiently accurate --- even if we additionally add laplace noise to preserve $\epsilon=0.1$-differential privacy.

\textbf{Count Sketch with Banlists} In our simulations we also consider an authentication server that bans the most popular $B=10^4$ passwords in a dataset to help flatten the password distribution and protect users against online attacks. Theoretical anaylsis indicates that directly banning the most popular passwords is the most effective way to increase the minimum entropy of the password distribution~\cite{BKPS:ACMEC13}. We remark that one additional benefit of using a Count Sketch datastructure is that it can be used to help implement this type of policy i.e., if a user attempts to register with password $pw$ and $\EstimateP{pw}{\sigma}$ is already too high then the user will be required to pick a different password~\cite{HTS:SchHerMit10}.

 We evaluate the performance of $\DALock$ in the presence of banlists. Recall that we let $\SampledData{\AllUser, B}$ denote the dataset $\SampledData{\AllUser}$ with the $B$ most common passwords removed following the normalized probabilities model of ~\cite{BKPS:ACMEC13} to model how affected users will update their passwords in response to the banlist. In particular, we assume users who are affected by the policy will pick a new passwords following the empirical distribution induced by $\SampledData{\AllUser, B}$. We then train the Count Sketch on the updated dataset i.e., $\sigma_{-B} = \text{Add}(\SampledData{\AllUser, B})$ as follows. 


\subsubsection{Frequency Oracle from Password Models}
As we previously discussed there are several reasons why an organization might prefer not to use a Count Sketch for frequency estimation e.g., privacy concerns or limited sample size. An alternative is to instantiate the frequency oracle with a password model. This could be a heuristic password strength meter, a more sophisticated model based on Neural Networks, Probabilistic Context Free Grammars or Markov Models or an empirical estimate based on Hashcat. The primary advantage to this approach is that the model can be deployed immediately even before an organization has any users and there are no privacy concerns. 

We adopted the $\ZXCVBN$ password strength meter~\cite{USENIX:Wheeler16} as prior empirical studies  demonstrate that it is one of the most accurate password strength meters \cite{CCS:GolDur18}. We used the Password Guessing Service \cite{USENIX:USBCCKKMMS15} to obtain guessing numbers for Neural Network, PCFG, Hashcat, and Markov Models ---  we also considered the  minimum guessing number across all four models as suggested in \cite{USENIX:USBCCKKMMS15}. For example, if a password $pw$ had guessing number $g$ we might estimate that $\EstP{pw_i} =1/g$. One challenge that we needed to address was that the estimates we obtain do not always yield a probability distribution e.g., for $\ZXCVBN$ we have $\sum_{i=1}^{10000}\EstP{pw_i} \gg 1$ where $i$ ranges over the top $10^4$ passwords in the dataset. Thus, before deploying the frequency estimator in $\DALock$ we renormalized our estimates so that $\sum_{i=1}^{10000}\EstP{pw_i} =1$. 


\subsection{Modeling the Attacker}\label{section:ExperimentDesign-subsection:SimulateAttacker} 
The final component of our simulation is a model of the attacker. We take a conservative approach and model an untargetted attacker with complete knowledge of the password distribution. Following Kerckhoff's principle we also assume that the attacker has access to the complete description of the $\DALock$ mechanism. In particular, for any password $pw$ we assume that the attacker knows both the true probability $ \TrueP{pw}$ and the estimated probability $\EstP{pw}$.  Finally, we also assume that the attacker is given the complete sequence of login times $t_1^u \leq t_2^u \leq  \ldots \leq 24 \times 180$ for each user $u$ over a 180 day time span as well as the outcome of each login attempt e.g., at time $t_i^u$ user $u$ will succeed after 2 incorrect guesses. 

{\bf Remark:} We conservatively aim to overestimate the capabilities of an untargetted online attacker. In practice, the online attacker will be able to able to approximate $ \TrueP{pw}$  and $\EstP{pw}$ overtime by interacting with the $\DALock$ server e.g., by setting up dummy accounts to test many times he can submit a particular incorrect guess without exceeding the hit count. Similarly, the attacker would not necessarily know the exact login times for a user, but this conservative assumption makes it feasible to precisely characterize the optimal behavior of an attacker. In practice, an online attacker might wait several days in between guesses to avoid accidently locking the user's account based on the number of consecutive incorrect login attempts. 

\subsubsection{Optimizing Attack Strategies} 
The goal of the attacker is to maximize the probability of cracking each password within the fixed $180$ day time span. For example, the attacker might try to find popular passwords $pw$ where the ratio $\EstP{pw}/\TrueP{pw}$ is small so that the increased hit count is smaller than intended. We formalize the attacker's optimal strategy in terms of the \textsf{Password Knapsack} problem $(\PK)$. Unsurprisingly, the password knapsack probelm turns out to be $\NP$-hard (as we prove in the appendix), but there are several heuristic algorithms the $\Adversary$ can use which yield nearly optimal strategies in practice. 

Recall that we assume that the $\Adversary$ has perfect knowledge of the distribution and probability estimates for each password $pw$. We also assume $\Adversary$ knows the $\DALock$ security parameters $\strikeThreshold$ and $\hitCountThreshold$. Furthermore, for each user $u$ we assume that the attacker is given the complete sequence of login times $t_1^u \leq t_2^u \leq  \ldots \leq 24 \times 180$ for each user $u$ over a 180 day time span as well as the outcome of each login attempt e.g., at time $t_i^u$ user $u$ will succeed after 2 incorrect guesses. In particular, at any point in time $t < 24\times 180$  the attacker can infer the current strike threshold and hit count threshold for any user $u$. We denote by  $\strikeThresholdOfU{u,t}$ (resp. $\hitCountThresholdOfU{u,t}$) the strike (resp. hit count) threshold  for user $u$ at time $t$ assuming that the attacker does not submit any of his own guesses. 

Supposing that the attacker wishes to avoid locking down the user's account before time $t$ the cummulative (estimated) probability of all guesses submitted before that time should be at most $\hitCountThresholdOfU{u,t}':=\hitCountThreshold- \hitCountThresholdOfU{u,t}$. Similarly, we let $M(t)$ denote the maximum number of gueses that the attacker can sneak in over the first $t$ hours without locking down the account i.e., because $\strikeThresholdOfU{u,t'}  \geq \strikeThreshold$ at some time $t' \leq t$. 
 
Fixing the time parameter $t$ the  attacker's goal is to find a subset $S_t \subseteq \AllPassword$ of $M(t)$ passwords to check such that 
\begin{equation} \label{eq:attackerConstraint}
\vspace{-0.2cm} 
\sum_{pw \in S_t} \EstP{pw} \leq \hitCountThresholdOfU{u,t}' \ . \vspace{-0.1cm} 
\end{equation}
After checking the passwords in $S_t$ the attacker can still check one more password $pw_{hold} \not\in S_t$ before the account is locked down. Given a set $S_t$ and a holdout password $pw_{hold} \not\in S_t$ the probability that the attacker succeeds is 
\begin{equation}\vspace{-0.2cm} \label{eq:attackerSuccess} \TrueP{pw_{hold}} + \sum_{pw \in S_t}\TrueP{pw} \ . \vspace{-0.1cm} \end{equation}

 Thus, the goal of the attacker is to find a subset $S_t$ of size $|S_t| \leq M(t)$ maximizing his success rate (eq \ref{eq:attackerSuccess}) subject to the constaint in  equation \ref{eq:attackerConstraint}.

 \mypara{Password Knapsack Problem}  Given a password dictionary \\$\{pw_1, \ldots, pw_n\}$ we formally define the \textsf{P}assword \textsf{K}napsack($\PK$) problem as the following integer program with indicator variables $s_i \in \{0,1\}$ and $l_i=\{0,1\}$ for each password $pw_i$. The attackers goal is to select a holdout password and a separate subset of $M$ ($=M(t)$) passwords with total `weight' (estimated probability) at most $\hitCountThreshold'$ ($= \hitCountThresholdOfU{u,t}'$) 
 
 $$
\begin{array}{crl}
	&\max {\displaystyle{\sum_i {(s_i + l_i) \cdot \TrueP{pw_{i}}}}} \\
	subject\ to, &\\
	&\sum_i{s_i \cdot \EstimateP{pw_i}{\sigma}) \le \hitCountThreshold'} \\
	&\sum_i s_i \le M\\
	&\sum_i l_i \le 1\\
	&\forall i~ l_i + s_i \le 1\\
	where,\\
	& \forall i, s_i, l_i \in \{0,1\}
\end{array}
$$
Intutively, setting $s_i$ = 1 means $pw_i$ is selected to be placed in the ``password knapsack" $S\subseteq \AllPassword$, i.e. to be used for dictionary attack. Setting $l_i=1$ indicates that password $pw_i$ is used as holdout password. This is equivalent to the following optimization problem. The constraints ensure that $|S| \leq M$ and we pick exactly one holdout password that is not already in $S$. 

\mypara{Solving the \textsf{P}assword \textsf{K}napsack} To maximize the number of cracked passwords an online attacker can compute $M(t)$ and $\hitCountThresholdOfU{u,t}':=\hitCountThreshold- \hitCountThresholdOfU{u,t}$ for each time $t \leq 24 \times 180$ and solve the corresponding \textsf{P}assword \textsf{K}napsack problem. Given optimal solutions $(pw_{hold,t}^*, S_t^*)$ for each time $t$ the attacker will pick the solution that maximizes the number of cracked passwords as in equation \ref{eq:attackerSuccess}. We remark that the calculations above need to be repeated for each different user $u$ since the values $M(t)$ and  $\hitCountThresholdOfU{u,t}'$ may vary due to different visitation schedules.

\mypara{Solving Password Knapsack}  Unfortunately, the \textsf{P}assword \textsf{K}napsack problem is $\NP$-hard as we prove in \textbf{Theorem}~\ref{appendix:ProofOfPasswordKnapsack} in the Appendix via a straightforward reduction from Subset Sum. In all of instances we considered we found that the optimal choice for the holdout password was simply $pw_1$ the most likely password in the distribution. Once we fix our holdout password our problem reduces to the two dimensional knapsack problem. We remark that $\PK$ can be viewed as a two dimensional knapsack problem. 

Assuming $P\neq NP$ the two dimensional knapsack probelm does not even admit a polynomial time approximation scheme ($\PTAS$) \cite{kulik2010there} in contrast to the regular knapsack problem which has fully polynomial time approximation scheme ($\FPTAS$)). Thus, we consider two heuristic approaches to solve the password knapsack problem:  $\mathsf{D}$antizig's $\mathsf{A}$lgorithm $\mathsf{B}$ased\cite{Dan:OR57} approach (\DAB) and $\mathsf{F}$easible $\mathsf{M}$ost $\mathsf{P}$romising $\mathsf{P}$assword $\mathsf{F}$irst approach(\FMPPF).

$\DAB$ (\textbf{Algorithm}~\ref{algorithm:Dantizig}, Appendix) sorts the remaining passwords $\mathcal{P}_{\tilde{\Pi}} = \{pw_2, \ldots\, pw_n\}$ based on the ratios $\frac{\TrueP{pw_i}}{\EstP{pw_i}}$ and select candidates based on the sorted order until we either select $M$ passwords or until selecting another password would exceed our capacity $\hitCountThreshold'$. $\FMPPF$ (\textbf{Algorithm}~\ref{algorithm:FMPPF}, Appendix) sorts the remaining passwords based on the true probability $\TrueP{pw_i}$  and simply selects password $pw$ in sorted order until we either select $M$ passwords or until selecting another password would exceed our capacity $\hitCountThreshold'$.  We discuss the advantages and disadvantages to both heuristics in the appendix. Intuitively, $\FMPPF$ (resp. $\DAB$) will perform better when $M$ (resp. $\hitCountThreshold'$) is the limiting constraint.

We found that $\FMPPF$ generally performs better than $\DAB$ despite of its simplicity. In addition, our simuation shows that $\FMPPF$'s performance is close to optimal. Practically speaking, one generally expect $\EstP{pw_i} \approx \TrueP{pw_i}$ especially when $pw_i$ is a popular password. In such case, $\DAB$ can hardly gain advantages from underestimation. Further more, imagine one bucket passwords by probability ranges, there are plenty of passwords in each bucket. Intuitively, picking passwords ordered by $\TrueP{pw_i}$ should produce an (almost) optimal solution (quickly). Thus, we choose to present the results based on $\FMPPF$ approach.

\section{Experimental Results}\label{section:experimentalresult} 
We empirically evaluated the performance of $\DALock$ under a variety of scenarios. During each simulation we had $10^6$ honest users register with an authentication server running $\DALock$ and login over a period of $180$ days. To analyze usability we ran the simulations without an online password attacker and measured  the unwanted lockout rate i.e., the fraction of user accounts that were locked due to honest mistakes. To analyze security we added an untargetted online attacker to the simulation and measured the fraction of user passwords that the attacker cracked.

The results of our simulations are summarized in \textbf{Figures}~\ref{figure:dictionaryAttack9.375},~\ref{figure:usability9.375}, ~\ref{figure:dictionaryAttack7.0}, ~\ref{figure:usability7.0},~\ref{figure:dictionaryAttackPrune}, and ~\ref{figure:usabilityPrune}\footnote{We included additional experimental results in \textbf{Appendix} ~\ref{appendix:experimentalResults} for interested readers. }.  The first four figures evaluate the security (\textbf{Figures}~\ref{figure:dictionaryAttack9.375} and~\ref{figure:dictionaryAttack7.0}) and usability (\textbf{Figures}~\ref{figure:usability9.375} and~\ref{figure:usability7.0}) in the absence of a banlist.  The last two figures evaluate the usability (\textbf{Figure}   ~\ref{figure:usabilityPrune} ) and security (\textbf{Figure} ~\ref{figure:dictionaryAttackPrune} ) of $\DALock$ when the authentication server bans the top $B=10^4$ passwords in our dataset. The X-axis of each plot represents the time span over 180 days. And the Y-axis represents percentage of compromised users (unwanted locked out rate) for security (usability) experiments.

\mypara{Implementation Details}  In each of our implementations $\DALock$ we instantiated $\strikeThreshold=10$ using hit count parameters $\hitCountThreshold \in \{ 2^{-7.0}, 2^{-9.375}\}$ (no banlist) and $\hitCountThreshold=2^{-11}$ (with banlist). In each batch of experiments we used one of our three password datsets $\SampledData{\AllUser}$ (RockYou, Yahoo!, or LinkedIn) to define our password distribution and we instantiated the frequency oracle using (1) a variety of password models including $\ZXCVBN$~\cite{USENIX:Wheeler16}, Hashcat, Markov Models, PCFGs and Neural Networks~\cite{USENIX:USBCCKKMMS15}\footnote{We relied on the Password Guessing Service to obtain cracking numbers for Hashcat, Markov Models, PCFGs and Neural Networks~\cite{USENIX:USBCCKKMMS15}. We found that the Neural Network model failed to crack many weak passwords in our datasets as it was configured not to guess short passwords. As such we did not directly use Neural Networks to implement our frequency oracle. However, the Neural Network guessing numbers are included in our Min-all frequency oracle which uses the minimum guessing number over all models.} and (2) an $\epsilon$-differentially private count sketch trained on a subsample $\SampledData{\AllUser_{r\%}}$ of containing $r\%$ of the original dataset $\SampledData{\AllUser}$ with $r \in \{1\%, 5\%, 10\%, \mathtt{All}\}$. When instantiated with banlist we used the dataset $\SampledData{\AllUser, B}$ instead of $\SampledData{\AllUser}$ i.e., we  removing the $B=10^4$ most common passwords.


\mypara{Baseline} We used the classical $3$-strike mechanism  and the $10$-strike mechanisms (recommend by Brostoff et. al \cite{brostoff2003ten} to improve usability) as a baseline for comparison. We remark that this is equivalent to $\KPsiDALock{3}{\hitCountThreshold=\infty}$ and $\KPsiDALock{10}{\hitCountThreshold=\infty}$ respectively. Our results suggest that one can improve {\em both} security and usability by replacing the classic 3-strike throttling mechanism with $(10,\hitCountThreshold)-\DALock$. Our results demonstrate that $\KPsiDALock{10}{\hitCountThreshold}$ greatly outperforms the classic $10$-strikes throttling mechanism without significant usability loss -- from a usability standpoint decreasing $\hitCountThreshold$ can only increase the unwanted lockout rate. We discuss our findings in more detail below.

\subsection{Usability}\label{section:ExperimentResult-usability} 
\textbf{Figures} ~\ref{figure:usability9.375} and ~\ref{figure:usability7.0} highlight the usability of $\DALock$ in the absence of a banlist with hit count parameters $\hitCountThreshold = 2^{-9.375}$ and $\hitCountThreshold=2^{-7.0}$. When $\hitCountThreshold=2^{-7.0}$ we find that $\DALock$ {\em always} outperforms the classical $3$-strikes mechanism regardless of how the frequency oracle is instantiated. When $\hitCountThreshold=2^{-9.375}$ $\DALock$ still outperforms the classical $3$-strikes mechanism when instantiated with a count-sketch --- even when we train on just $r=1\%$ of the dataset and add laplace noise to achieve $\epsilon=0.1$-differential privacy. When we instantiate $\DALock$ password models the results were mixed e.g., $\ZXCVBN$ and Hashcat had superior usability while Markov Models and Probabilistic Context Free Grammars performed poorly. 

\textbf{Figure} \ref{figure:usabilityPrune} highlights the usability benefit of banning the top $10^4$ passwords. While a few users might be inconvenienced during the password registration our simulations indicate that the unwanted lockout rates for $\DALock$ are greatly reduced even when we adopt a stricter $\hitCountThreshold = 2^{-11}$. Intuitively, the banlist allows us to avoid locking out users who select and overly popular password as one of their five alternate passwords for other accounts. We remark that when we instantiate $\DALock$ with a count sketch that the usability results are virtually identical to the $10$-strikes policy --- even if we use a $\epsilon=0.1$ -differentially private count sketch trained on just $1\%$ of the data.


\mypara{10-strike Mechanism is user friendly} Brostoff et. al \cite{brostoff2003ten} proposed that one should replace 3-strike mechanism with 10-strike mechanism to achieve higher usability. Our simulation results clearly align with their recommendations. Based on our plots, 10-strike mechanism results in unwanted locked out rate close to zero. However, deploying this mechanism also has a high security cost as indicated by our simulations.


\begin{figure*}\label{key1}
	\includegraphics[width=\linewidth, height = 4.5cm]{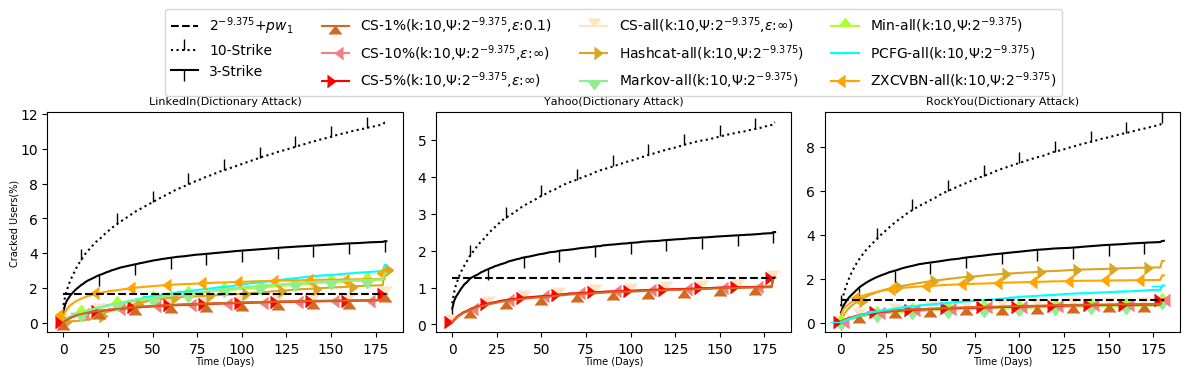}
	\caption{Security Measurement of $\DALock$ - $\hitCountThreshold = 2^{-9.375}$ }\label{figure:dictionaryAttack9.375}
	\includegraphics[width=\linewidth, height = 4.5cm]{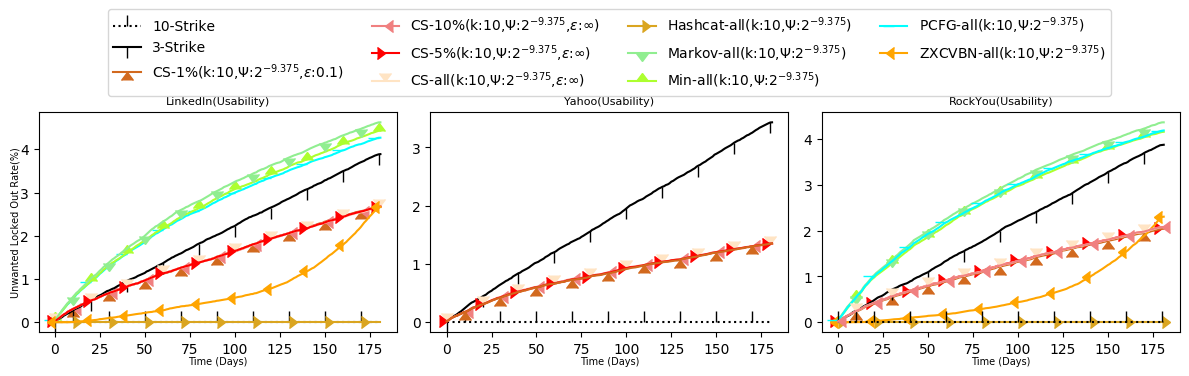}
	\caption{Usability Measurement of $\DALock$ - $\hitCountThreshold = 2^{-9.375}$ }\label{figure:usability9.375}
	\includegraphics[width=\linewidth, height = 4.5cm]{{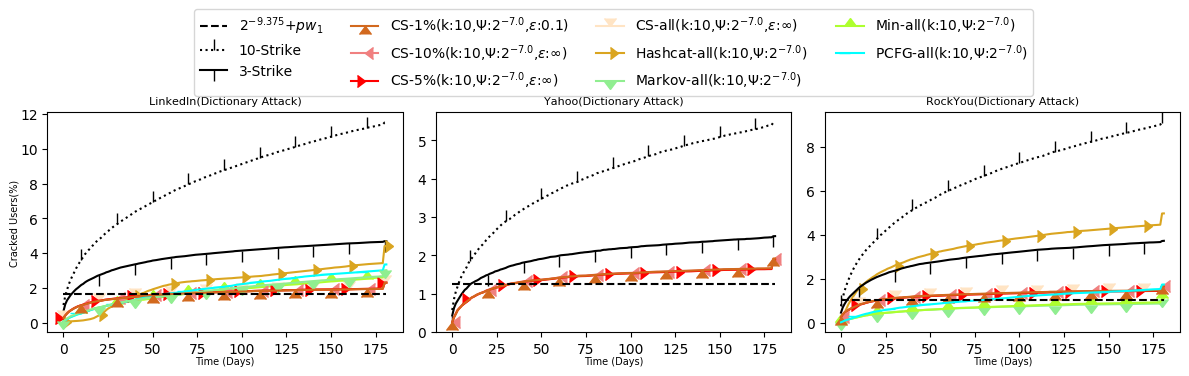}}
	\caption{Security Measurement of $\DALock$ - $\hitCountThreshold = 2^{-7.0}$ }\label{figure:dictionaryAttack7.0}
	\includegraphics[width=\linewidth, height = 4.5cm]{{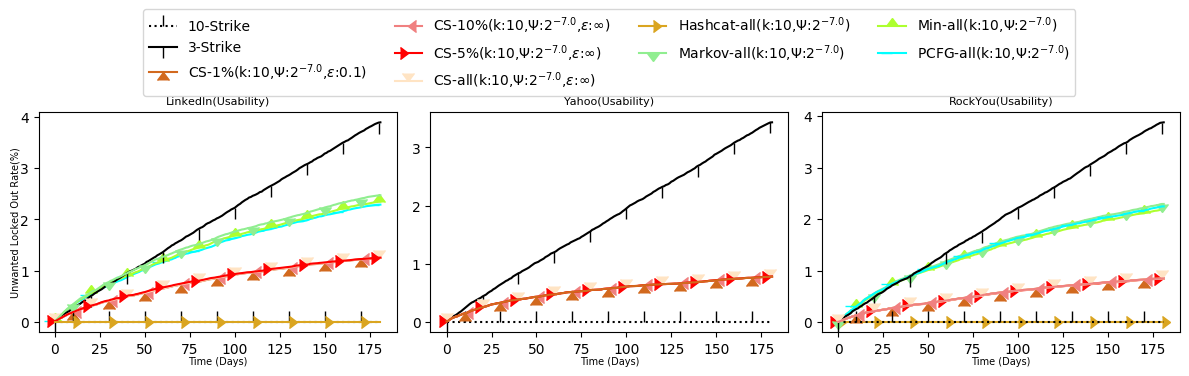}}
	\caption{Usability Measurement of $\DALock$ - $\hitCountThreshold = 2^{-7.0}$ }\label{figure:usability7.0}
\end{figure*}

\begin{figure*}\label{key2}
	\includegraphics[width=\linewidth, height = 4.5cm]{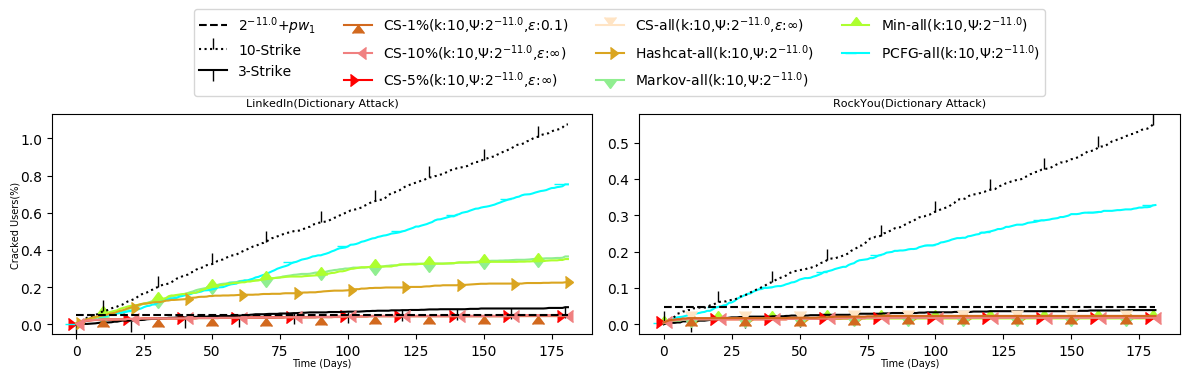}
	\vspace{-0.8cm}
	\caption{Security Measurement of $\DALock$ - $\hitCountThreshold = 2^{-11}$ (Banning top $B=10^4$ passwords)  }\label{figure:dictionaryAttackPrune}
	\includegraphics[width=\linewidth, height = 4.5cm]{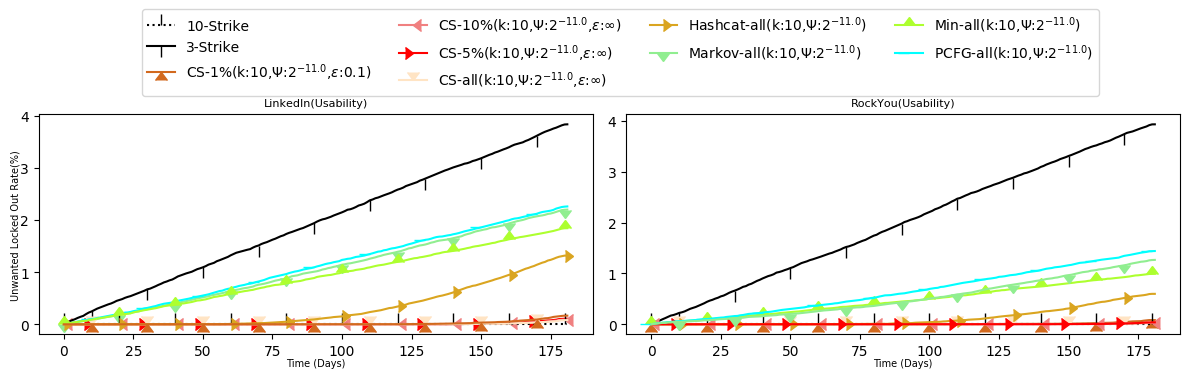}
		\vspace{-0.8cm}
	\caption{Usability Measurement of $\DALock$ - $\hitCountThreshold = 2^{-11}$ (Banning top $B=10^4$ passwords)}\label{figure:usabilityPrune}
		\vspace{-0.6cm}
\end{figure*}

\subsection{Security Results} \label{section:ExperimentResult-security} 
\textbf{Figures}~\ref{figure:dictionaryAttack9.375} and ~\ref{figure:dictionaryAttack7.0} evaluate the security of $\DALock$ in the absence of a banlist with hit count parameters $\hitCountThreshold = 2^{-9.375}$ and $\hitCountThreshold=2^{-7}$ respectively. For reference we also plotted the line $\hitCountThreshold + \TrueP(pw_1)$ which constitutes a theoretical upper bound on the attacker success rate when $\DALock$ is instantiated with a perfect frequency oracle. We found  that $\DALock$ always outperforms even the stricter $\strikeThreshold=3$-strikes policy under {\em all} insantiations of the frequency oracle (excluding Hashcat with $\hitCountThreshold=2^{-7}$). For example, the attacker cracked  roughly 4\% of users accounts when facing 3-strike mechanism (LinkedIn + RockYou) compared with $2\%$ when deploying  $\DALock$  with a differentially private count sketch.

\textbf{Figure}~\ref{figure:dictionaryAttackPrune} highlights the advantage of banning the most popular $B=10^4$ passwords. We remark that $\DALock$ always outperformed the $\strikeThreshold=10$-strikes policy. We also found that the $\strikeThreshold=3$-strikes policy and $\DALock$ (with a differentially private Count Sketch) were both {\em highly effective} at protecting user accounts with a compromised rate is about close to zero. ( $\approx$ 0.092\% for 3-strike and  $\approx$ 0.048\% for differential private Count Sketch).

\subsection{Summary and Discussion}\label{sec:experiment_summary}
We find that $\KPsiDALock{10}{\Psi}$ offers a superior security/usability tradeoff to the classical $\strikeThreshold$-strikes mechanism. Our experiments also highlight the security {\em and} usability benefits of banning overly popular passwords. We found that $\DALock$ can be reasonably instantiated with password strength models such as $\ZXCVBN$, Markov Models, Probabilistic Context Free Grammars and Neural Networks. However, we obtain the best security/usability tradeoffs when we ban the most popular passwords and when we  instantiate the $\DALock$ frequency oracle    with a differentially private count sketch. We found that the count sketch can be reliable trained from a smaller subsample containing just $1\%$ of the dataset, and even if we add enough noise to preserve $\epsilon-0.1$-differential privacy (strong privacy). This is promising news for a smaller organization that is considering  deploying $\DALock$.


\mypara{Limitations}\label{sec: Limitations}
Our empirical security results are all based on simulations. While we aim to model the authentication server, users and an attacker there will inevitably be some differences between the simulated/real-world behavior of the attacker/users. We also remark that our simulations do not model the behavior of targetted attackers. Extending $\DALock$ to protect against targetted attackers is an important research question that is beyond the scope of the current paper. Finally, we remark that larger organizations might distribute the workload across multiple authentication servers. In this case maintaining a synchronized state $(\strikeThresholdOfU{u}, \hitCountThresholdOfU{u})$ for each user $u$ could be challenging. To address this challenge it may be necessary to define a relaxation of our $\DALock$ mechanism where the states $(\strikeThresholdOfU{u}, \hitCountThresholdOfU{u})$ on each authentication server are not always assumed to be perfectly synchronized.

\section{Conclusion}
We present a novel {\em distribution aware} password throttling mechanism $\DALock$ that penalizes incorrect passwords proportionally to their popularity. We show that $\DALock$ can be reliably instantiated with either a password strength model such as $\ZXCVBN$ or with a differentially private count sketch. Our empirical analysis demonstrates that $\DALock$ offers a superior balance between security and usability and is particularly effective when used  in combination with a short banlist of overly popular passwords. For example, we are able to  reduce the success rate of an attacker to $0.05\%$ (compared to $1\%$ for the $10$-strikes mechanism) whilst simultaneously reducing the unwanted lockout rate to just $0.08\%$ (compared to $4\%$ for the $3$-strikes mechanism).  
\clearpage
\appendix
\section{$\DALock$}

\subsection{$\DALock$ Authentication Algorithm}
We supplement the pseudo code of $\DALock$ in this section to help readers understand how to implement $\DALock$ for authencation. The authencation process takes four arguments: username $u$, input password $pw$, salt $s_u$, and password popularity estimator $\sigma$. Before verifying the correctness of entered password, $\DALock$ first check if $u$'s account has already been locked or not based on $\hitCountThreshold_{u}$ and $\strikeThreshold{u}$. If the account is not locked, $\DALock$ proceeds and verify the correctness of the passwords. If the password is valid, $\DALock$ resets strike threshold $\strikeThresholdOfU{u}$ and grant user the access to the service. If the entered password is wrong then in addition to denying the access to the service, the server also increases $\hitCountThresholdOfU{u}$ and $\strikeThreshold$ by $\EstP{pw}$ and 1 respectively.

\begin{algorithm}[!htb]
	\caption{\textbf{$\DALock$}: Novel Password Distribution Aware Throttling Mechanism }\label{algorithm:DALock}
	\begin{algorithmic}[1]
		\Function{login}{$u$, $pw_u$, $\sigma$,$s_u$} 
		\If {$\hitCountThresholdOfU{u} \geq \hitCountThreshold$ or  $\strikeThresholdOfU{u} \geq \strikeThreshold$ }
		\State Reject Login
		\EndIf
		\If{$hash(pw,s_u)$ == $hash(pw_u,s_u)$}
		\State Reset $\strikeThresholdOfU{u}$ to $\strikeThreshold$ 
		\State Grant Access
		\Else
		\State $\hitCountThresholdOfU{u} \leftarrow \hitCountThresholdOfU{u} + \EstimateP{pw}{\sigma}$
		\State $\strikeThresholdOfU{u} \leftarrow \strikeThresholdOfU{u}$ + 1
		\State Deny Access
		\EndIf
		\EndFunction
	\end{algorithmic}
\end{algorithm} 
\subsection{Password Knapsack is $\NP$ hard}

\begin{theorem}[Hardness of Password Knapsack]\label{appendix:ProofOfPasswordKnapsack}
	Find optimal solution for password knapsack is $\NP$-hard.
\end{theorem}

\mypara{Proof:}
We first formally define subset sum problem, and then prove password knapsack is $\NP$ hard by showing the reduction from subset sum to it.
\begin{definition}[Subset Sum]
	Given Partition instance $x_1,\ldots,x_{n} \in (0,2^m]$ and target sum value $T$. The  goal is to find $S \subseteq [n]$ s.t. $\sum_{i\in S} x_i = T$? 
\end{definition}
\textbf{Reduction}: One can create the following password knapsack instance 
\begin{itemize}
	\item Set $\gamma = \sum_{i=1}^n x_i$,
	\item Set $\psi = T/(2\gamma )< \frac{1}{2}$,
	\item Set $CS(p_i)= f(p_{i}) = x_i/(2\gamma)$ for $i=1,\ldots, n$
	\item Set $f(p_{last}) = 1-\sum_{i =1}^{n} p_i = 1/2 > \psi$. 
\end{itemize}
If $S$ exists for partition instance then attacker can use $S$ for password knapsack to crack $p_{last}+T/(2\gamma)$ passwords. On the other hand let $S$ be the optimal password knapsack solution such that $\sum_{i \in S} CS(p_i) \leq \psi$ then the attacker cracks at most $p_{last}+\sum_{i \in S} f(p_i) \leq 1/2 + \psi$ passwords. If equality holds then $\sum_{i \in S} f(p_i) = \psi$ which implies $\sum_{i \in S} x_i = T$ by definition of $\psi$.

\subsection{Solving $\PK$ with Heruistics}
In this section we supplement the missing details of algorithm $\DAB$ and $\FMPPF$ mentioned in \textbf{Section}~\ref{section:ExperimentDesign-subsection:SimulateAttacker}.  

The $\DAB$ approach takes three inputs: a sorted password dictionary based on the ratio of actual popularity and estimated popularity $\frac{\EstP{pw}}{\TrueP{pw}}$: $\mathcal{P}_{\tilde{\Pi}} = \{pw_{\tilde{\Pi}(1)}, \ldots, pw_{\tilde{\Pi}(n)} \}$, attack budget $\Psi$ and K. The algorithm keeps placing passwords into the knapsack S based on the sorted order until it cannot further add some password $pw$. At this points, $\DAB$ compares $\TrueP{pw}$ with $\TrueP{S}$ and sets S to be the one with higher values. After that, the algorithm repeat the above process by scanning throught the whole dictionary. At the end, since only $K$ passwords is allowed to be used for guessing, the algorithm returns $K$ passwords based on their actual probability

Primary incentives of using this algorithm are 1) to take advantage of underestimated passwords and 2) to avoid (severely) overestimated ones. There are several drawbacks of $\DAB$. Firstly, the progress can be slow because priority are given to significantly underestimated passwords. Intuitively, for popular passwords ($\TrueP{pw}$ large) the ratio $\frac{\TrueP{pw}}{\EstP{pw}}$ is likely to be close to 1, therefore, attempts with popular ones are likely to be delayed. Secondly, unlike vanilla version of Knaspack, $\DAB$ may not yield a 2-approximation due to the additional constraint on the number of passwords one can place in the Knapsack. Third, computation cost is slightly higher for running $\DAB$ though both algorithms terminate reasonably quickly. 

\begin{algorithm}[!htb]
	\caption{\textbf{$\DAB$ Attack }}\label{algorithm:Dantizig}
	\textbf{Return:}  An array of password sorted in the order of guessing
	\begin{algorithmic}[1]
		\Function{$\DAB$ Attack}{$\mathcal{P}_{\tilde{\Pi}}, \Psi, M(T)$}
		\State $S$= []
		\While{$S$ changes}
		\For{ $pw \in \mathcal{P}_{\tilde{\Pi}}$  }
		\If{$\EstP{pw}$ $> \Psi$ }
		\State \textbf{continue}
		\EndIf
		\If{$\EstP{S\cup pw}$ $< \Psi$ and $|S| \le$ $M(T)$  }
		\State $S \leftarrow S \cup pw$ 
		\ElsIf{$\TrueP{pw} > \TrueP{S}$ }
		\State $S \leftarrow$ \{p\}
		\EndIf
		\EndFor
		\EndWhile
		\State \Return Top $M(T)$ passwords from $S$ based on actual popularity.
		\EndFunction
	\end{algorithmic}
\end{algorithm}

An alternative to $\DAB$ is $\FMPPF$. It takes three input parameters $\mathcal{P} = pw_1, \ldots, pw_{n} \}$, attack budget $\Psi$ and $M(T)$ and selects passwords greedily. $\FMPPF$ differs from $\DAB$ in the following two aspects. Firstly, $\FMPPF$ uses a password dictioary sorted based on the actual popularity only, which can be easily obtained in reaf life. Secondly, to save computational cost $\FMPPF$  terminates once it finds $K$ passwords that are suitable for attacks and stop further explore the dictionary. The pseudo can be found in \textbf{Algorithm}~\ref{algorithm:FMPPF}

In short time attack scenarios, $\FMPPF$ offers better chance of success than $\DAB$ by attempting popular ones first. For long term case, $\FMPPF$ should still be able to achieve almost optimal results given an abundant choice of passwords.  In fact, based on the empirical results (in \textbf{section}~\ref{section:ExperimentResult-security}), the performance of $\FMPPF$ is very close to theoretical upper bounds ($\Psi + \TrueP{pw_1}$ ). 

\begin{algorithm}[!htb]
	\caption{\textbf{$\FMPPF$ Approach}}\label{algorithm:FMPPF}
	\textbf{Return:} An array of password sorted in the order of guessing
	\begin{algorithmic}[1]
		\Function{\FMPPF}{$\mathcal{P}, \Psi, M(T)$}
		\State $S$= []
		\For{ $pw \in \mathcal{P}$  }
		\If{$\EstP{S \cup p}$ $< \Psi$ and $|S| < M(T)$  }
		\State $S \leftarrow S \cup pw$ 
		\EndIf
		\EndFor
		\State \Return S
		\EndFunction
	\end{algorithmic}
\end{algorithm}

\section{Simulating User's Mistakes}
In this section we elaborate the details for simulating users' mistakes missing in \textbf{Section}~\ref{section:ExperimentDesign-subsection:SimulateUser}. To help reader visualize the process of simulating mistakes, we plot the flowchart in \textbf{figure}~\ref{figure:flowChartTypo}. The starting point is to simulate recall errors. Based on the empirical results of existing literatures\cite{CCS:CWPCR17,SP:CAAJR16}, we set the probability of makeing an recall error to 2.4\%. Recall that when we generate user's profile, each user has five ``passwords from other services". Therefore, we simulate recall error by choose one of them to be the \emph{password user intends to enter}. On top of this process, we further simulate typos (on the password intended to enter) with probability $\approx 5\%$. Condition on making typos, we simulate this step by choosing a type of typos with their conditional probability summarized in \textbf{Table}~\ref{Table:TypoTypes}.

\begin{figure}[htb]
	\begin{center}
		\includegraphics[height=2in,width=\linewidth]{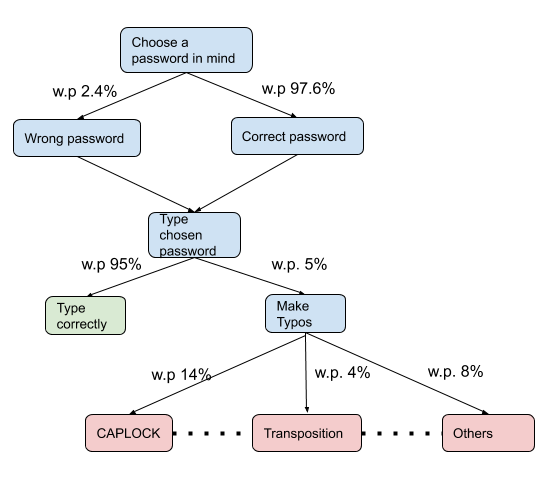}
		\caption{Flow Chart for Simulating Users' mistake}\label{figure:flowChartTypo}
	\end{center}
\end{figure}

\begin{table}[h]
	\begin{tabular}{|c|c|}
		\hline
		Typo Types           & Chance of Mistake(Rounded \%) \\ \hline
		CapLock On           & 14                            \\ \hline
		Shift First Char     & 4                             \\ \hline
		One Extra Insertion  & 12                            \\ \hline
		One Extra Deletion   & 12                            \\ \hline
		One Char Replacement & 31                            \\ \hline
		Transposition        & 4                             \\ \hline
		Two Deletions         & 3                             \\ \hline
		Two Insertions        & 3                             \\ \hline
		Two Replacements      & 10                            \\ \hline
		Others               & 8                             \\ \hline
	\end{tabular}
	\caption{Typo Distributions\cite{CCS:CWPCR17}}
	\label{Table:TypoTypes}
	\vspace{-1cm}
\end{table}

\section{More Experimental Results}\label{appendix:experimentalResults}
We provide more experimental results for curious readers to better understand the performance of $\DALock$ in multiple scenarios. To clearly demonstrates the impact of $\hitCountThreshold$ on security and usability we choose to visualize the following set of parameters $\{2^{-6}, 2^{-7},2^{-8},2^{-9}, 2^{-10}\}$(\textbf{Figure}~\ref{figure:appendix_attacker} and ~\ref{figure:appendix_usability}). When one set $\hitCountThreshold$ to an over conservative, e.g. $2^{-10}$ or smaller, dictionary attackers are not able to crack a significant potion of users accounst; however, usebility can be a concern as infrequent passwords burns $\hitCountThreshold_{u}$ quickly.(see curve CS-all($k:10,\hitCountThreshold:2^{-10},\epsilon$) across all figures). Another extreme approach is adopting aggresively large $\hitCountThreshold$, e.g. $2^{-6}$. Based on the plots(curve CS-all($k:10,\hitCountThreshold:2^{-6},\epsilon$)), the usability performance is satisfactory while security risk is enlarged (but still better than 3-strike). Beyond standard Count Sketch, we supplement more results on applying differential privacy to Count-Sketch(\textbf{Figure}~\ref{figure:appendix_attacker_dp} and  ~\ref{figure:appendix_usability_dp}). We discovered that differential privacy hardly have any negative impact on the performance of $\DALock$ especially when the dataset is large. Further more, we supplement more results on subsampling to show $\sigma_{r\%}$ is effective for a wide range of $\hitCountThreshold$ even when the sampling rate is 1\%.(\textbf{Figure}~\ref{figure:appendix_attacker_sample} and  ~\ref{figure:appendix_usability_sample}) Finally, we present the results of differentially private $\DALock$ trained on subsampling dataset $\SampledData{U_{1\%}}$. Despite the datasets become 100 times smaller, $\DP{\sigma_{1\%}}$ still performs reasonly well.(\textbf{Figure}~\ref{figure:appendix_attacker_dpandsample} and ~\ref{figure:appendix_usability_dpandsample}).

\mypara{Almost Optimal Heruistic}  If one assumes that $\TrueP{pw} \approx \EstP{pw}$ for popular passwords, then theoretically $\Adversary$ can compromise at most $\hitCountThreshold + \TrueP{pw_1}$ users accounts by using $pw_1$ as holdout passwords. Based on our simulation results, we found that $\Adversary$ is very close to achieve such threshold by adopting $\FMPPF$ algorithm metioned in \textbf{Section}~\ref{section:ExperimentDesign-subsection:SimulateAttacker}. To help readers identify this property, we highlight the upperbouds a series of upperbounds(e.g. $2^{-6} + \TrueP{pw_1}$ etc)

\clearpage
\begin{figure*}[htb]
	\includegraphics[width=\linewidth,height=4cm]{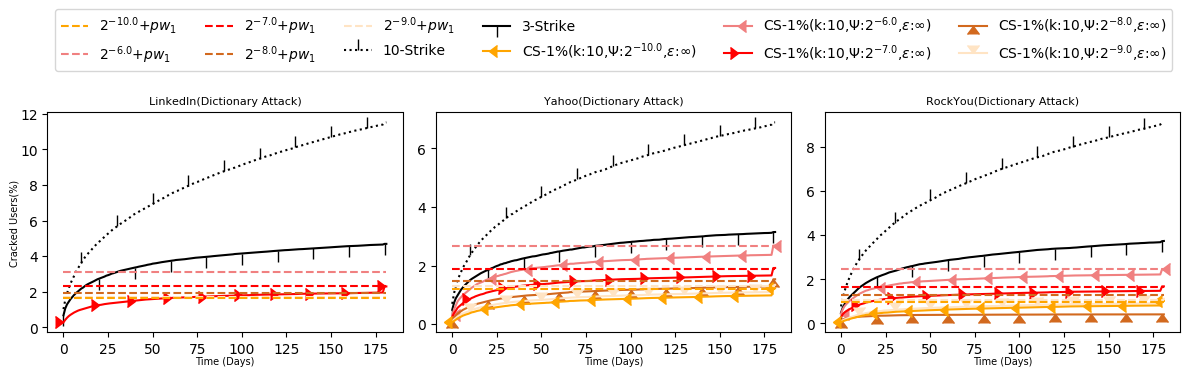}
	\caption{Security: $\sigma$}\label{figure:appendix_attacker}
	\includegraphics[width=\linewidth,height=4cm]{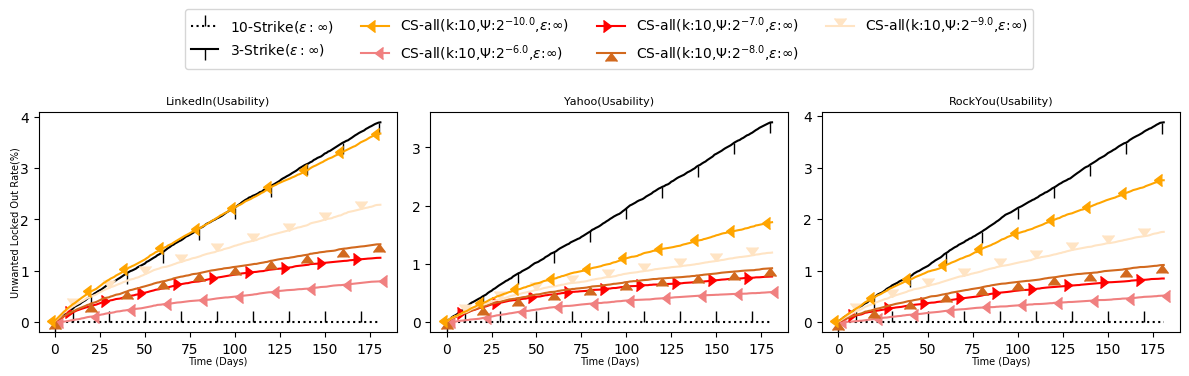}
	\caption{Usability:  $\sigma$}\label{figure:appendix_usability}
	\includegraphics[width=\linewidth,height=4cm]{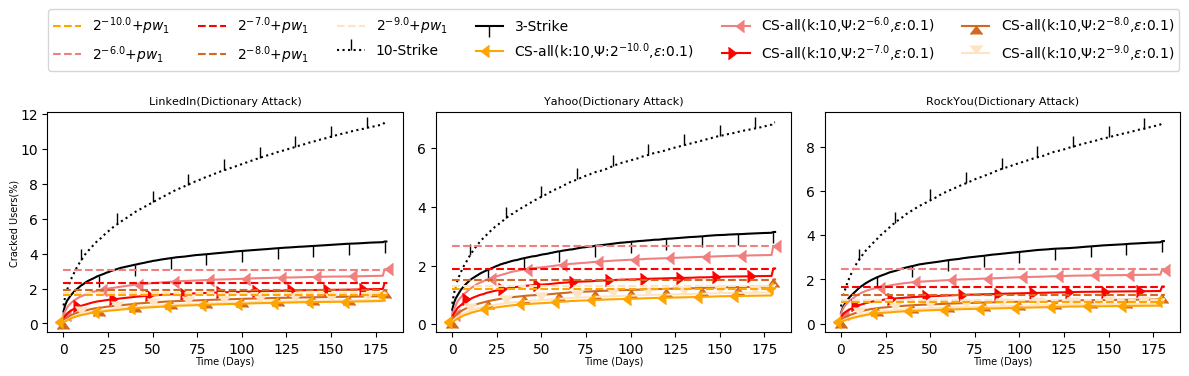}
	\caption{Security: $\DP{0.1}{\sigma}$}\label{figure:appendix_attacker_dp}
	\includegraphics[width=\linewidth,height=4cm]{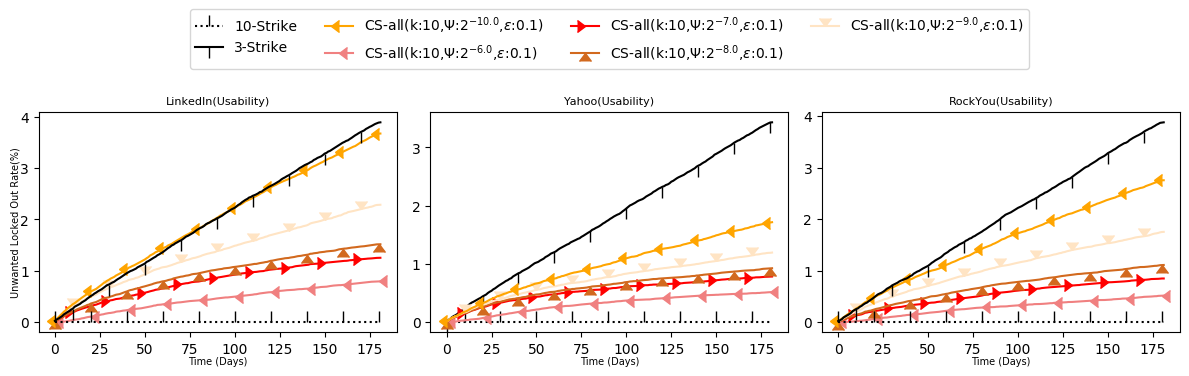} 
	\caption{Usability: $\DP{0.1}{\sigma}$}\label{figure:appendix_usability_dp}
\end{figure*}
\begin{figure*}[htb]
			\includegraphics[width=\linewidth,height=4cm]{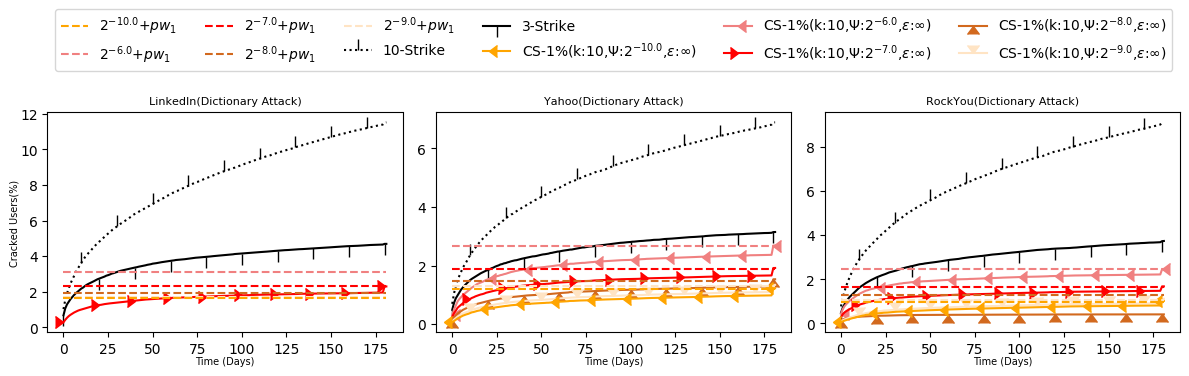}
	\caption{Security: $\sigma_{1\%}$}\label{figure:appendix_attacker_sample}
	\includegraphics[width=\linewidth,height=4cm]{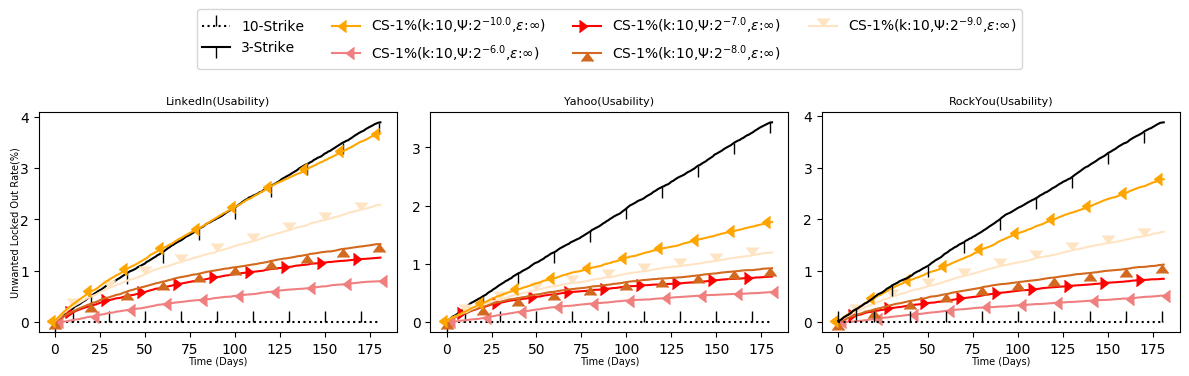}
	\caption{Usability: $\sigma_{1\%}$}\label{figure:appendix_usability_sample}
	\includegraphics[width=\linewidth,height=4cm]{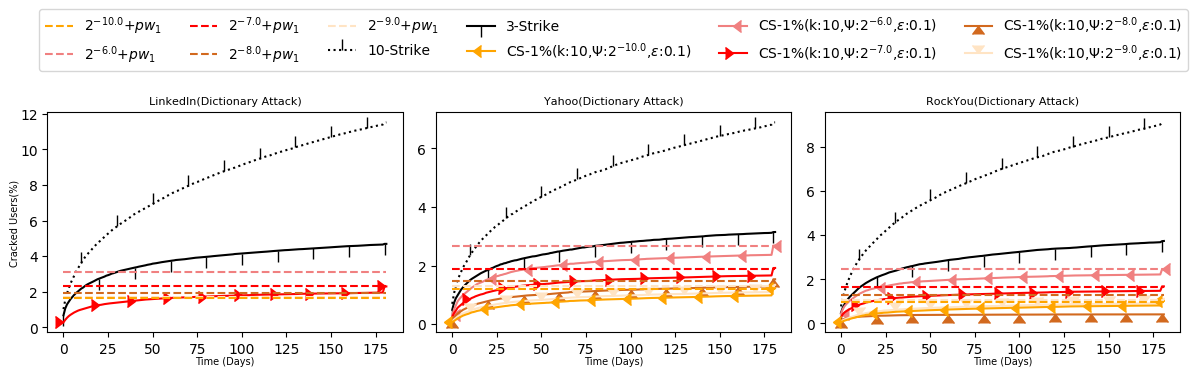}
	\caption{Security: $\DP{0.1}{\sigma_{1\%}}$ }\label{figure:appendix_attacker_dpandsample}
	\includegraphics[width=\linewidth,height=4cm]{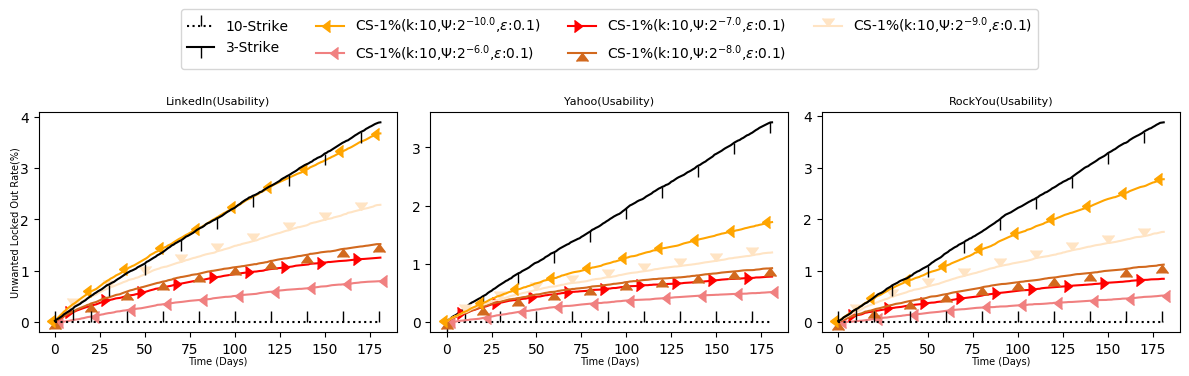} 
	\caption{Usability:$\DP{0.1}{\sigma_{1\%}}$}\label{figure:appendix_usability_dpandsample}
	\end{figure*}
\clearpage




\begin{acks}
This research is supported by the Purdue Research Foundation. A preliminary draft of this paper was presented at the WAY 2019 workshop\footnote{https://wayworkshop.org}. The authors thank the WAY PC members for valuable feedback.
\end{acks}
 
\newpage
\bibliographystyle{ACM-Reference-Format}

\begin{thebibliography}{00}


\ifx \showCODEN    \undefined \def \showCODEN     #1{\unskip}     \fi
\ifx \showDOI      \undefined \def \showDOI       #1{#1}\fi
\ifx \showISBNx    \undefined \def \showISBNx     #1{\unskip}     \fi
\ifx \showISBNxiii \undefined \def \showISBNxiii  #1{\unskip}     \fi
\ifx \showISSN     \undefined \def \showISSN      #1{\unskip}     \fi
\ifx \showLCCN     \undefined \def \showLCCN      #1{\unskip}     \fi
\ifx \shownote     \undefined \def \shownote      #1{#1}          \fi
\ifx \showarticletitle \undefined \def \showarticletitle #1{#1}   \fi
\ifx \showURL      \undefined \def \showURL       {\relax}        \fi
\providecommand\bibfield[2]{#2}
\providecommand\bibinfo[2]{#2}
\providecommand\natexlab[1]{#1}
\providecommand\showeprint[2][]{arXiv:#2}

\bibitem[\protect\citeauthoryear{Blocki, Blum, and Datta}{Blocki
  et~al\mbox{.}}{2013}]%
        {AC:BloBluDat13}
\bibfield{author}{\bibinfo{person}{Jeremiah Blocki}, \bibinfo{person}{Manuel
  Blum}, {and} \bibinfo{person}{Anupam Datta}.}
  \bibinfo{year}{2013}\natexlab{}.
\newblock \showarticletitle{Naturally Rehearsing Passwords}.
  \bibinfo{pages}{361--380}.
\newblock
\showDOI{%
\url{https://doi.org/10.1007/978-3-642-42045-0_19}}


\bibitem[\protect\citeauthoryear{Blocki, Datta, and Bonneau}{Blocki
  et~al\mbox{.}}{2016}]%
        {NDSS:BloDatBon16}
\bibfield{author}{\bibinfo{person}{Jeremiah Blocki}, \bibinfo{person}{Anupam
  Datta}, {and} \bibinfo{person}{Joseph Bonneau}.}
  \bibinfo{year}{2016}\natexlab{}.
\newblock \showarticletitle{Differentially Private Password Frequency Lists}.
\newblock


\bibitem[\protect\citeauthoryear{Blocki, Harsha, and Zhou}{Blocki
  et~al\mbox{.}}{2018}]%
        {SP:BloHarZho18}
\bibfield{author}{\bibinfo{person}{Jeremiah Blocki}, \bibinfo{person}{Benjamin
  Harsha}, {and} \bibinfo{person}{Samson Zhou}.}
  \bibinfo{year}{2018}\natexlab{}.
\newblock \showarticletitle{On the Economics of Offline Password Cracking}.
  \bibinfo{pages}{853--871}.
\newblock
\showDOI{%
\url{https://doi.org/10.1109/SP.2018.00009}}


\bibitem[\protect\citeauthoryear{Blocki, Komanduri, Procaccia, and
  Sheffet}{Blocki et~al\mbox{.}}{2013}]%
        {BKPS:ACMEC13}
\bibfield{author}{\bibinfo{person}{Jeremiah Blocki}, \bibinfo{person}{Saranga
  Komanduri}, \bibinfo{person}{Ariel Procaccia}, {and} \bibinfo{person}{Or
  Sheffet}.} \bibinfo{year}{2013}\natexlab{}.
\newblock \showarticletitle{Optimizing password composition policies}. In
  \bibinfo{booktitle}{{\em Proceedings of the fourteenth ACM conference on
  Electronic commerce}}. ACM, \bibinfo{pages}{105--122}.
\newblock


\bibitem[\protect\citeauthoryear{Bonneau}{Bonneau}{2012}]%
        {SP:Bonneau12}
\bibfield{author}{\bibinfo{person}{Joseph Bonneau}.}
  \bibinfo{year}{2012}\natexlab{}.
\newblock \showarticletitle{The Science of Guessing: Analyzing an Anonymized
  Corpus of 70 Million Passwords}. \bibinfo{pages}{538--552}.
\newblock
\showDOI{%
\url{https://doi.org/10.1109/SP.2012.49}}


\bibitem[\protect\citeauthoryear{Bonneau, Herley, {van Oorschot}, and
  Stajano}{Bonneau et~al\mbox{.}}{2012}]%
        {SP:BHVS12}
\bibfield{author}{\bibinfo{person}{Joseph Bonneau}, \bibinfo{person}{Cormac
  Herley}, \bibinfo{person}{Paul~C. {van Oorschot}}, {and}
  \bibinfo{person}{Frank Stajano}.} \bibinfo{year}{2012}\natexlab{}.
\newblock \showarticletitle{The Quest to Replace Passwords: A Framework for
  Comparative Evaluation of Web Authentication Schemes}.
  \bibinfo{pages}{553--567}.
\newblock
\showDOI{%
\url{https://doi.org/10.1109/SP.2012.44}}


\bibitem[\protect\citeauthoryear{Brostoff and Sasse}{Brostoff and
  Sasse}{2003}]%
        {brostoff2003ten}
\bibfield{author}{\bibinfo{person}{Sacha Brostoff} {and}
  \bibinfo{person}{Angela Sasse}.} \bibinfo{year}{2003}\natexlab{}.
\newblock \showarticletitle{Ten strikes and you're out: Increasing the number
  of login attempts can improve password usability}.
\newblock  (\bibinfo{date}{07} \bibinfo{year}{2003}).
\newblock


\bibitem[\protect\citeauthoryear{Bursztein, Bethard, Fabry, Mitchell, and
  Jurafsky}{Bursztein et~al\mbox{.}}{2010}]%
        {SP:BBFNJ10}
\bibfield{author}{\bibinfo{person}{Elie Bursztein}, \bibinfo{person}{Steven
  Bethard}, \bibinfo{person}{Celine Fabry}, \bibinfo{person}{John~C. Mitchell},
  {and} \bibinfo{person}{Daniel Jurafsky}.} \bibinfo{year}{2010}\natexlab{}.
\newblock \showarticletitle{How Good Are Humans at Solving {CAPTCHAs}? {A}
  Large Scale Evaluation}. \bibinfo{pages}{399--413}.
\newblock
\showDOI{%
\url{https://doi.org/10.1109/SP.2010.31}}


\bibitem[\protect\citeauthoryear{Bursztein, Martin, and Mitchell}{Bursztein
  et~al\mbox{.}}{2011}]%
        {CCS:BurMarMit11}
\bibfield{author}{\bibinfo{person}{Elie Bursztein}, \bibinfo{person}{Matthieu
  Martin}, {and} \bibinfo{person}{John~C. Mitchell}.}
  \bibinfo{year}{2011}\natexlab{}.
\newblock \showarticletitle{Text-based {CAPTCHA} strengths and weaknesses}.
  \bibinfo{pages}{125--138}.
\newblock
\showDOI{%
\url{https://doi.org/10.1145/2046707.2046724}}


\bibitem[\protect\citeauthoryear{Charikar, Chen, and {Farach-Colton}}{Charikar
  et~al\mbox{.}}{2002}]%
        {ICALP:ChaCheFar02}
\bibfield{author}{\bibinfo{person}{Moses Charikar}, \bibinfo{person}{Kevin~C.
  Chen}, {and} \bibinfo{person}{Martin {Farach-Colton}}.}
  \bibinfo{year}{2002}\natexlab{}.
\newblock \showarticletitle{Finding Frequent Items in Data Streams}.
  \bibinfo{pages}{693--703}.
\newblock
\showDOI{%
\url{https://doi.org/10.1007/3-540-45465-9_59}}


\bibitem[\protect\citeauthoryear{Chatterjee, Athayle, Akhawe, Juels, and
  Ristenpart}{Chatterjee et~al\mbox{.}}{2016}]%
        {SP:CAAJR16}
\bibfield{author}{\bibinfo{person}{Rahul Chatterjee}, \bibinfo{person}{Anish
  Athayle}, \bibinfo{person}{Devdatta Akhawe}, \bibinfo{person}{Ari Juels},
  {and} \bibinfo{person}{Thomas Ristenpart}.} \bibinfo{year}{2016}\natexlab{}.
\newblock \showarticletitle{{pASSWORD} {tYPOS} and How to Correct Them
  Securely}. \bibinfo{pages}{799--818}.
\newblock
\showDOI{%
\url{https://doi.org/10.1109/SP.2016.53}}


\bibitem[\protect\citeauthoryear{Chatterjee, Woodage, Pnueli, Chowdhury, and
  Ristenpart}{Chatterjee et~al\mbox{.}}{2017}]%
        {CCS:CWPCR17}
\bibfield{author}{\bibinfo{person}{Rahul Chatterjee}, \bibinfo{person}{Joanne
  Woodage}, \bibinfo{person}{Yuval Pnueli}, \bibinfo{person}{Anusha Chowdhury},
  {and} \bibinfo{person}{Thomas Ristenpart}.} \bibinfo{year}{2017}\natexlab{}.
\newblock \showarticletitle{The {TypTop} System: Personalized Typo-Tolerant
  Password Checking}. \bibinfo{pages}{329--346}.
\newblock
\showDOI{%
\url{https://doi.org/10.1145/3133956.3134000}}


\bibitem[\protect\citeauthoryear{Cormode and Muthukrishnan}{Cormode and
  Muthukrishnan}{2005}]%
        {JoA:CorMut05}
\bibfield{author}{\bibinfo{person}{Graham Cormode} {and} \bibinfo{person}{Shan
  Muthukrishnan}.} \bibinfo{year}{2005}\natexlab{}.
\newblock \showarticletitle{An improved data stream summary: the count-min
  sketch and its applications}.
\newblock \bibinfo{journal}{{\em Journal of Algorithms\/}}
  \bibinfo{volume}{55}, \bibinfo{number}{1} (\bibinfo{year}{2005}),
  \bibinfo{pages}{58--75}.
\newblock


\bibitem[\protect\citeauthoryear{Dantzig}{Dantzig}{1957}]%
        {Dan:OR57}
\bibfield{author}{\bibinfo{person}{George~B Dantzig}.}
  \bibinfo{year}{1957}\natexlab{}.
\newblock \showarticletitle{Discrete-variable extremum problems}.
\newblock \bibinfo{journal}{{\em Operations research\/}} \bibinfo{volume}{5},
  \bibinfo{number}{2} (\bibinfo{year}{1957}), \bibinfo{pages}{266--288}.
\newblock


\bibitem[\protect\citeauthoryear{Dwork}{Dwork}{2011}]%
        {ECS:Dwork11}
\bibfield{author}{\bibinfo{person}{Cynthia Dwork}.}
  \bibinfo{year}{2011}\natexlab{}.
\newblock \showarticletitle{Differential privacy}.
\newblock \bibinfo{journal}{{\em Encyclopedia of Cryptography and Security\/}}
  (\bibinfo{year}{2011}), \bibinfo{pages}{338--340}.
\newblock


\bibitem[\protect\citeauthoryear{Erlingsson, Pihur, and Korolova}{Erlingsson
  et~al\mbox{.}}{2014}]%
        {CCS:ErlPihKor14}
\bibfield{author}{\bibinfo{person}{{\'U}lfar Erlingsson},
  \bibinfo{person}{Vasyl Pihur}, {and} \bibinfo{person}{Aleksandra Korolova}.}
  \bibinfo{year}{2014}\natexlab{}.
\newblock \showarticletitle{{RAPPOR}: Randomized Aggregatable
  Privacy-Preserving Ordinal Response}. \bibinfo{pages}{1054--1067}.
\newblock
\showDOI{%
\url{https://doi.org/10.1145/2660267.2660348}}


\bibitem[\protect\citeauthoryear{Florencio and Herley}{Florencio and
  Herley}{2007}]%
        {FloHer:WWW07}
\bibfield{author}{\bibinfo{person}{Dinei Florencio} {and}
  \bibinfo{person}{Cormac Herley}.} \bibinfo{year}{2007}\natexlab{}.
\newblock \showarticletitle{A large-scale study of web password habits}. In
  \bibinfo{booktitle}{{\em Proceedings of the 16th international conference on
  World Wide Web}}. ACM, \bibinfo{pages}{657--666}.
\newblock


\bibitem[\protect\citeauthoryear{Freeman, Jain, D{\"u}rmuth, Biggio, and
  Giacinto}{Freeman et~al\mbox{.}}{2016}]%
        {NDSS:FJDBG16}
\bibfield{author}{\bibinfo{person}{David Freeman}, \bibinfo{person}{Sakshi
  Jain}, \bibinfo{person}{Markus D{\"u}rmuth}, \bibinfo{person}{Battista
  Biggio}, {and} \bibinfo{person}{Giorgio Giacinto}.}
  \bibinfo{year}{2016}\natexlab{}.
\newblock \showarticletitle{Who Are You? {A} Statistical Approach to Measuring
  User Authenticity}.
\newblock


\bibitem[\protect\citeauthoryear{Gao, Yan, Cao, Zhang, Lei, Tang, Zhang, Zhou,
  Wang, and Li}{Gao et~al\mbox{.}}{2016}]%
        {NDSS:GYCZLT16}
\bibfield{author}{\bibinfo{person}{Haichang Gao}, \bibinfo{person}{Jeff Yan},
  \bibinfo{person}{Fang Cao}, \bibinfo{person}{Zhengya Zhang},
  \bibinfo{person}{Lei Lei}, \bibinfo{person}{Mengyun Tang},
  \bibinfo{person}{Ping Zhang}, \bibinfo{person}{Xin Zhou},
  \bibinfo{person}{Xuqin Wang}, {and} \bibinfo{person}{Jiawei Li}.}
  \bibinfo{year}{2016}\natexlab{}.
\newblock \showarticletitle{A Simple Generic Attack on Text Captchas}.
\newblock


\bibitem[\protect\citeauthoryear{ghacks}{ghacks}{2011}]%
        {News:AmazonTypo}
ghacks \bibinfo{year}{2011}\natexlab{}.
\newblock \bibinfo{title}{Amazon Login May Accept Password Variants}.
\newblock   (\bibinfo{year}{2011}).
\newblock
\showURL{%
\url{https://www.ghacks.net/2011/01/31/amazon-login-may-accept-password-variants/}}


\bibitem[\protect\citeauthoryear{Golla, Bailey, and D{\"u}rmuth}{Golla
  et~al\mbox{.}}{2017}]%
        {SOUPS:GBD17}
\bibfield{author}{\bibinfo{person}{Maximilian Golla}, \bibinfo{person}{Daniel~V
  Bailey}, {and} \bibinfo{person}{Markus D{\"u}rmuth}.}
  \bibinfo{year}{2017}\natexlab{}.
\newblock \showarticletitle{" I want my money back!" Limiting Online
  Password-Guessing Financially.}. In \bibinfo{booktitle}{{\em SOUPS}}.
\newblock


\bibitem[\protect\citeauthoryear{Golla and D{\"u}rmuth}{Golla and
  D{\"u}rmuth}{2018}]%
        {CCS:GolDur18}
\bibfield{author}{\bibinfo{person}{Maximilian Golla} {and}
  \bibinfo{person}{Markus D{\"u}rmuth}.} \bibinfo{year}{2018}\natexlab{}.
\newblock \showarticletitle{On the Accuracy of Password Strength Meters}.
  \bibinfo{pages}{1567--1582}.
\newblock
\showDOI{%
\url{https://doi.org/10.1145/3243734.3243769}}


\bibitem[\protect\citeauthoryear{Gordon and Lundeen}{Gordon and
  Lundeen}{2014}]%
        {gordon2014efficiently}
\bibfield{author}{\bibinfo{person}{Ariel Gordon} {and}
  \bibinfo{person}{Richard~Allen Lundeen}.} \bibinfo{year}{2014}\natexlab{}.
\newblock \bibinfo{title}{Efficiently throttling user authentication}.
\newblock   (\bibinfo{date}{Nov.~25} \bibinfo{year}{2014}).
\newblock
\newblock
\shownote{US Patent 8,898,752.}


\bibitem[\protect\citeauthoryear{Harsha, Morton, Blocki, Springer, and
  Dark}{Harsha et~al\mbox{.}}{2020}]%
        {harsha2020bicycle}
\bibfield{author}{\bibinfo{person}{Benjamin Harsha}, \bibinfo{person}{Robert
  Morton}, \bibinfo{person}{Jeremiah Blocki}, \bibinfo{person}{John Springer},
  {and} \bibinfo{person}{Melissa Dark}.} \bibinfo{year}{2020}\natexlab{}.
\newblock \showarticletitle{Bicycle Attacks Considered Harmful: Quantifying the
  Damage of Widespread Password Length Leakage}.
\newblock \bibinfo{journal}{{\em arXiv preprint arXiv:2002.01513\/}}
  (\bibinfo{year}{2020}).
\newblock


\bibitem[\protect\citeauthoryear{Have I Been Pwned}{Have I Been Pwned}{2019}]%
        {WebSite:HaveIBeenPwned}
Have I Been Pwned \bibinfo{year}{2019}\natexlab{}.
\newblock \bibinfo{title}{Have I Been Pwned}.
\newblock   (\bibinfo{year}{2019}).
\newblock
\showURL{%
\url{https://haveibeenpwned.com}}


\bibitem[\protect\citeauthoryear{{Herley} and {Van Oorschot}}{{Herley} and {Van
  Oorschot}}{2012}]%
        {PasswordPersistence}
\bibfield{author}{\bibinfo{person}{C. {Herley}} {and} \bibinfo{person}{P. {Van
  Oorschot}}.} \bibinfo{year}{2012}\natexlab{}.
\newblock \showarticletitle{A Research Agenda Acknowledging the Persistence of
  Passwords}.
\newblock \bibinfo{journal}{{\em IEEE Security Privacy\/}}
  \bibinfo{volume}{10}, \bibinfo{number}{1} (\bibinfo{date}{Jan}
  \bibinfo{year}{2012}), \bibinfo{pages}{28--36}.
\newblock
\showISSN{1558-4046}
\showDOI{%
\url{https://doi.org/10.1109/MSP.2011.150}}


\bibitem[\protect\citeauthoryear{Inc}{Inc}{}]%
        {AppleDP}
\bibfield{author}{\bibinfo{person}{Apple Inc}.}
\newblock \bibinfo{title}{Apple Differential Privacy Technical Overview}.
\newblock   (\bibinfo{year}{????}).
\newblock
\showURL{%
\url{https://www.apple.com/privacy/docs/Differential_Privacy_Overview.pdf}}
\newblock
\shownote{Retrieved 25, Apr. 2019.}


\bibitem[\protect\citeauthoryear{Kogan, Manohar, and Boneh}{Kogan
  et~al\mbox{.}}{2017}]%
        {CCS:KogManBon17}
\bibfield{author}{\bibinfo{person}{Dmitry Kogan}, \bibinfo{person}{Nathan
  Manohar}, {and} \bibinfo{person}{Dan Boneh}.}
  \bibinfo{year}{2017}\natexlab{}.
\newblock \showarticletitle{T/Key: Second-Factor Authentication From Secure
  Hash Chains}. \bibinfo{pages}{983--999}.
\newblock
\showDOI{%
\url{https://doi.org/10.1145/3133956.3133989}}


\bibitem[\protect\citeauthoryear{Komanduri, Shay, Kelley, Mazurek, Bauer,
  Christin, Cranor, and Egelman}{Komanduri et~al\mbox{.}}{2011}]%
        {KSKMBCCE:SIGCHI11}
\bibfield{author}{\bibinfo{person}{Saranga Komanduri}, \bibinfo{person}{Richard
  Shay}, \bibinfo{person}{Patrick~Gage Kelley}, \bibinfo{person}{Michelle~L
  Mazurek}, \bibinfo{person}{Lujo Bauer}, \bibinfo{person}{Nicolas Christin},
  \bibinfo{person}{Lorrie~Faith Cranor}, {and} \bibinfo{person}{Serge
  Egelman}.} \bibinfo{year}{2011}\natexlab{}.
\newblock \showarticletitle{Of passwords and people: measuring the effect of
  password-composition policies}. In \bibinfo{booktitle}{{\em Proceedings of
  the SIGCHI Conference on Human Factors in Computing Systems}}. ACM,
  \bibinfo{pages}{2595--2604}.
\newblock


\bibitem[\protect\citeauthoryear{Kulik and Shachnai}{Kulik and
  Shachnai}{2010}]%
        {kulik2010there}
\bibfield{author}{\bibinfo{person}{Ariel Kulik} {and} \bibinfo{person}{Hadas
  Shachnai}.} \bibinfo{year}{2010}\natexlab{}.
\newblock \showarticletitle{There is no EPTAS for two-dimensional knapsack}.
\newblock \bibinfo{journal}{{\it Inform. Process. Lett.}}
  \bibinfo{volume}{110}, \bibinfo{number}{16} (\bibinfo{year}{2010}),
  \bibinfo{pages}{707--710}.
\newblock


\bibitem[\protect\citeauthoryear{LinkedIn}{LinkedIn}{nd}]%
        {Dataset:LinkedIn}
LinkedIn \bibinfo{year}{n.d.}\natexlab{}.
\newblock \bibinfo{title}{LinkedIn Password Corpus}.
\newblock   (\bibinfo{year}{n.d.}).
\newblock
\showURL{%
\url{https://hashes.org/public.php}}


\bibitem[\protect\citeauthoryear{Malone and Maher}{Malone and Maher}{2012}]%
        {DavKev:WWW12}
\bibfield{author}{\bibinfo{person}{David Malone} {and} \bibinfo{person}{Kevin
  Maher}.} \bibinfo{year}{2012}\natexlab{}.
\newblock \showarticletitle{Investigating the distribution of password
  choices}. In \bibinfo{booktitle}{{\em Proceedings of the 21st international
  conference on World Wide Web}}. ACM, \bibinfo{pages}{301--310}.
\newblock


\bibitem[\protect\citeauthoryear{Naor, Pinkas, and Ronen}{Naor
  et~al\mbox{.}}{2019}]%
        {CCS:NaoPinRon19}
\bibfield{author}{\bibinfo{person}{Moni Naor}, \bibinfo{person}{Benny Pinkas},
  {and} \bibinfo{person}{Eyal Ronen}.} \bibinfo{year}{2019}\natexlab{}.
\newblock \showarticletitle{How to (not) Share a Password: Privacy Preserving
  Protocols for Finding Heavy Hitters with Adversarial Behavior}.
  \bibinfo{pages}{1369--1386}.
\newblock
\showDOI{%
\url{https://doi.org/10.1145/3319535.3363204}}


\bibitem[\protect\citeauthoryear{Narayanan and Shmatikov}{Narayanan and
  Shmatikov}{2006}]%
        {arXiv:NarShm06}
\bibfield{author}{\bibinfo{person}{Arvind Narayanan} {and}
  \bibinfo{person}{Vitaly Shmatikov}.} \bibinfo{year}{2006}\natexlab{}.
\newblock \showarticletitle{How to break anonymity of the netflix prize
  dataset}.
\newblock \bibinfo{journal}{{\em arXiv preprint cs/0610105\/}}
  (\bibinfo{year}{2006}).
\newblock


\bibitem[\protect\citeauthoryear{Narayanan and Shmatikov}{Narayanan and
  Shmatikov}{2008}]%
        {UTA:NarShm08}
\bibfield{author}{\bibinfo{person}{Arvind Narayanan} {and}
  \bibinfo{person}{Vitaly Shmatikov}.} \bibinfo{year}{2008}\natexlab{}.
\newblock \showarticletitle{Robust de-anonymization of large datasets (how to
  break anonymity of the Netflix prize dataset)}.
\newblock \bibinfo{journal}{{\em University of Texas at Austin\/}}
  (\bibinfo{year}{2008}).
\newblock


\bibitem[\protect\citeauthoryear{Pham}{Pham}{2019}]%
        {DuoWeakPassword}
\bibfield{author}{\bibinfo{person}{Thu Pham}.} \bibinfo{year}{2019}\natexlab{}.
\newblock \bibinfo{title}{STOP THE PWNAGE: 81\% OF HACKING INCIDENTS USED
  STOLEN OR WEAK PASSWORDS}.
\newblock   (\bibinfo{year}{2019}).
\newblock
\showURL{%
\url{https://duo.com/decipher/stop-the-pwnage-81-of-hacking-incidents-used-stolen-or-weak-passwords}}
\newblock
\shownote{Retrieved \today.}


\bibitem[\protect\citeauthoryear{Pinkas and Sander}{Pinkas and Sander}{2002}]%
        {CCS:PinSan02}
\bibfield{author}{\bibinfo{person}{Benny Pinkas} {and} \bibinfo{person}{Tomas
  Sander}.} \bibinfo{year}{2002}\natexlab{}.
\newblock \showarticletitle{Securing Passwords Against Dictionary Attacks}.
  \bibinfo{pages}{161--170}.
\newblock
\showDOI{%
\url{https://doi.org/10.1145/586110.586133}}


\bibitem[\protect\citeauthoryear{prowebscraper.}{prowebscraper.}{2019}]%
        {captchaSolver}
prowebscraper. \bibinfo{year}{2019}\natexlab{}.
\newblock \bibinfo{title}{Top 10 Captcha Solving Services Compared}.
\newblock   (\bibinfo{year}{2019}).
\newblock
\showURL{%
\url{https://prowebscraper.com/blog/top-10-captcha-solving-services-compared/}}


\bibitem[\protect\citeauthoryear{RockYou}{RockYou}{2010}]%
        {Dataset:RockYou}
RockYou \bibinfo{year}{2010}\natexlab{}.
\newblock \bibinfo{title}{RockYou Password Corpus}.
\newblock   (\bibinfo{year}{2010}).
\newblock
\showURL{%
\url{http://downloads.skullsecurity.org/passwords/rockyou.txt.bz2.}}


\bibitem[\protect\citeauthoryear{Sandhu, Desa, and Ganesan}{Sandhu
  et~al\mbox{.}}{2005}]%
        {sandhu2005system}
\bibfield{author}{\bibinfo{person}{Ravi Sandhu}, \bibinfo{person}{Colin Desa},
  {and} \bibinfo{person}{Karuna Ganesan}.} \bibinfo{year}{2005}\natexlab{}.
\newblock \bibinfo{title}{System and method for password throttling}.
\newblock   (\bibinfo{date}{April~19} \bibinfo{year}{2005}).
\newblock
\newblock
\shownote{US Patent 6,883,095.}


\bibitem[\protect\citeauthoryear{Schechter, Herley, and Mitzenmacher}{Schechter
  et~al\mbox{.}}{2010}]%
        {HTS:SchHerMit10}
\bibfield{author}{\bibinfo{person}{Stuart Schechter}, \bibinfo{person}{Cormac
  Herley}, {and} \bibinfo{person}{Michael Mitzenmacher}.}
  \bibinfo{year}{2010}\natexlab{}.
\newblock \showarticletitle{Popularity is everything: A new approach to
  protecting passwords from statistical-guessing attacks}. In
  \bibinfo{booktitle}{{\em Proceedings of the 5th USENIX conference on Hot
  topics in security}}. USENIX Association, \bibinfo{pages}{1--8}.
\newblock


\bibitem[\protect\citeauthoryear{Schecter and Herley}{Schecter and
  Herley}{2016}]%
        {SchHer:MSR18}
\bibfield{author}{\bibinfo{person}{S Schecter} {and} \bibinfo{person}{C
  Herley}.} \bibinfo{year}{2016}\natexlab{}.
\newblock \showarticletitle{The Binomial Ladder Frequency Filter and its
  Applications to Shared Secrets}.
\newblock \bibinfo{journal}{{\em MSR-TR-2018-18\/}} (\bibinfo{year}{2016}).
\newblock


\bibitem[\protect\citeauthoryear{Team}{Team}{}]%
        {AppleDPTeam}
\bibfield{author}{\bibinfo{person}{Apple Differential~Privacy Team}.}
\newblock \bibinfo{title}{Learning with Privacy at Scale}.
\newblock   (\bibinfo{year}{????}).
\newblock
\showURL{%
\url{https://machinelearning.apple.com/2017/12/06/learning-with-privacy-at-scale.html}}
\newblock
\shownote{Retrieved 25, Apr. 2019.}


\bibitem[\protect\citeauthoryear{TechNewsWorld}{TechNewsWorld}{2019}]%
        {DictionaryAttack:Microsoft}
TechNewsWorld \bibinfo{year}{2019}\natexlab{}.
\newblock \bibinfo{title}{Microsoft Exposes Russian Cyberattacks on Phones,
  Printers, Video Decoders}.
\newblock   (\bibinfo{year}{2019}).
\newblock
\showURL{%
\url{https://www.technewsworld.com/story/86171.html}}


\bibitem[\protect\citeauthoryear{Tuerk}{Tuerk}{2019}]%
        {WebSite:GooglePasswordCheckUp}
\bibfield{author}{\bibinfo{person}{Andreas Tuerk}.}
  \bibinfo{year}{2019}\natexlab{}.
\newblock \bibinfo{title}{To stay secure online, Password Checkup has your
  back}.
\newblock   (\bibinfo{year}{2019}).
\newblock
\showURL{%
\url{https://www.blog.google/technology/safety-security/password-checkup/}}


\bibitem[\protect\citeauthoryear{Tung}{Tung}{2019}]%
        {DictionaryAttack:Ransomware}
\bibfield{author}{\bibinfo{person}{Liam Tung}.}
  \bibinfo{year}{2019}\natexlab{}.
\newblock \bibinfo{title}{Ransomware crooks hit Synology NAS devices with
  brute-force password attacks}.
\newblock   (\bibinfo{year}{2019}).
\newblock
\showURL{%
\url{https://www.zdnet.com/article/ransomware-crooks-hit-synology-nas-devices-with-brute-force-password-attacks/}}
\newblock
\shownote{Retrieved \today.}


\bibitem[\protect\citeauthoryear{Ur, Segreti, Bauer, Christin, Cranor,
  Komanduri, Kurilova, Mazurek, Melicher, and Shay}{Ur et~al\mbox{.}}{2015}]%
        {USENIX:USBCCKKMMS15}
\bibfield{author}{\bibinfo{person}{Blase Ur}, \bibinfo{person}{Sean~M.
  Segreti}, \bibinfo{person}{Lujo Bauer}, \bibinfo{person}{Nicolas Christin},
  \bibinfo{person}{Lorrie~Faith Cranor}, \bibinfo{person}{Saranga Komanduri},
  \bibinfo{person}{Darya Kurilova}, \bibinfo{person}{Michelle~L. Mazurek},
  \bibinfo{person}{William Melicher}, {and} \bibinfo{person}{Richard Shay}.}
  \bibinfo{year}{2015}\natexlab{}.
\newblock \showarticletitle{Measuring Real-World Accuracies and Biases in
  Modeling Password Guessability}. \bibinfo{pages}{463--481}.
\newblock


\bibitem[\protect\citeauthoryear{{von Ahn}, Blum, Hopper, and Langford}{{von
  Ahn} et~al\mbox{.}}{2003}]%
        {EC:vBHL03}
\bibfield{author}{\bibinfo{person}{Luis {von Ahn}}, \bibinfo{person}{Manuel
  Blum}, \bibinfo{person}{Nicholas~J. Hopper}, {and} \bibinfo{person}{John
  Langford}.} \bibinfo{year}{2003}\natexlab{}.
\newblock \showarticletitle{{CAPTCHA}: Using Hard {AI} Problems for Security}.
  \bibinfo{pages}{294--311}.
\newblock
\showDOI{%
\url{https://doi.org/10.1007/3-540-39200-9_18}}


\bibitem[\protect\citeauthoryear{Von~Ahn, Maurer, McMillen, Abraham, and
  Blum}{Von~Ahn et~al\mbox{.}}{2008}]%
        {von2008recaptcha}
\bibfield{author}{\bibinfo{person}{Luis Von~Ahn}, \bibinfo{person}{Benjamin
  Maurer}, \bibinfo{person}{Colin McMillen}, \bibinfo{person}{David Abraham},
  {and} \bibinfo{person}{Manuel Blum}.} \bibinfo{year}{2008}\natexlab{}.
\newblock \showarticletitle{recaptcha: Human-based character recognition via
  web security measures}.
\newblock \bibinfo{journal}{{\em Science\/}} \bibinfo{volume}{321},
  \bibinfo{number}{5895} (\bibinfo{year}{2008}), \bibinfo{pages}{1465--1468}.
\newblock


\bibitem[\protect\citeauthoryear{Wang, Cheng, Wang, Huang, and Jian}{Wang
  et~al\mbox{.}}{2017}]%
        {TIFS17:WCWPXG}
\bibfield{author}{\bibinfo{person}{Ding Wang}, \bibinfo{person}{Haibo Cheng},
  \bibinfo{person}{Ping Wang}, \bibinfo{person}{Xinyi Huang}, {and}
  \bibinfo{person}{Gaopeng Jian}.} \bibinfo{year}{2017}\natexlab{}.
\newblock \showarticletitle{Zipf’s law in passwords}.
\newblock \bibinfo{journal}{{\em IEEE Transactions on Information Forensics and
  Security\/}} \bibinfo{volume}{12}, \bibinfo{number}{11}
  (\bibinfo{year}{2017}), \bibinfo{pages}{2776--2791}.
\newblock


\bibitem[\protect\citeauthoryear{Wang, Jian, Huang, and Wang}{Wang
  et~al\mbox{.}}{2014}]%
        {EPRINT:WJHW14}
\bibfield{author}{\bibinfo{person}{Ding Wang}, \bibinfo{person}{Gaopeng Jian},
  \bibinfo{person}{Xinyi Huang}, {and} \bibinfo{person}{Ping Wang}.}
  \bibinfo{year}{2014}\natexlab{}.
\newblock \bibinfo{title}{Zipf's Law in Passwords}.
\newblock \bibinfo{howpublished}{Cryptology ePrint Archive, Report 2014/631}.
  (\bibinfo{year}{2014}).
\newblock
\newblock
\shownote{\url{http://eprint.iacr.org/2014/631}.}


\bibitem[\protect\citeauthoryear{Wang and Wang}{Wang and Wang}{2016}]%
        {ESORICS:WanWan16}
\bibfield{author}{\bibinfo{person}{Ding Wang} {and} \bibinfo{person}{Ping
  Wang}.} \bibinfo{year}{2016}\natexlab{}.
\newblock \showarticletitle{On the Implications of {Zipf}'s Law in Passwords}.
  \bibinfo{pages}{111--131}.
\newblock
\showDOI{%
\url{https://doi.org/10.1007/978-3-319-45744-4_6}}


\bibitem[\protect\citeauthoryear{Wang, Zhang, Wang, Yan, and Huang}{Wang
  et~al\mbox{.}}{2016}]%
        {CCS:WZWYH16}
\bibfield{author}{\bibinfo{person}{Ding Wang}, \bibinfo{person}{Zijian Zhang},
  \bibinfo{person}{Ping Wang}, \bibinfo{person}{Jeff Yan}, {and}
  \bibinfo{person}{Xinyi Huang}.} \bibinfo{year}{2016}\natexlab{}.
\newblock \showarticletitle{Targeted Online Password Guessing: An
  Underestimated Threat}. \bibinfo{pages}{1242--1254}.
\newblock
\showDOI{%
\url{https://doi.org/10.1145/2976749.2978339}}


\bibitem[\protect\citeauthoryear{Wheeler}{Wheeler}{2016}]%
        {USENIX:Wheeler16}
\bibfield{author}{\bibinfo{person}{Daniel~Lowe Wheeler}.}
  \bibinfo{year}{2016}\natexlab{}.
\newblock \showarticletitle{zxcvbn: Low-Budget Password Strength Estimation}.
  \bibinfo{pages}{157--173}.
\newblock


\bibitem[\protect\citeauthoryear{Ye, Tang, Fang, Zhu, Feng, Xu, Chen, and
  Wang}{Ye et~al\mbox{.}}{2018}]%
        {CCS:YTFZFX18}
\bibfield{author}{\bibinfo{person}{Guixin Ye}, \bibinfo{person}{Zhanyong Tang},
  \bibinfo{person}{Dingyi Fang}, \bibinfo{person}{Zhanxing Zhu},
  \bibinfo{person}{Yansong Feng}, \bibinfo{person}{Pengfei Xu},
  \bibinfo{person}{Xiaojiang Chen}, {and} \bibinfo{person}{Zheng Wang}.}
  \bibinfo{year}{2018}\natexlab{}.
\newblock \showarticletitle{Yet Another Text Captcha Solver: {A} Generative
  Adversarial Network Based Approach}. \bibinfo{pages}{332--348}.
\newblock
\showDOI{%
\url{https://doi.org/10.1145/3243734.3243754}}


\bibitem[\protect\citeauthoryear{Yuan~Tian}{Yuan~Tian}{2019}]%
        {EuroSP:THS19}
\bibfield{author}{\bibinfo{person}{Stuart~Schechter Yuan~Tian, Cormac~Herley}.}
  \bibinfo{year}{2019}\natexlab{}.
\newblock \showarticletitle{StopGuessing: Using Guessed Passwords to Thwart
  Online Guessing}. In \bibinfo{booktitle}{{\em 4th IEEE European Symposium on
  Security and Privacy}}. IEEE.
\newblock


\bibitem[\protect\citeauthoryear{ZDNet}{ZDNet}{2019}]%
        {News:FacebookCaseSensitiveNews}
ZDNet \bibinfo{year}{2019}\natexlab{}.
\newblock \bibinfo{title}{Facebook passwords are not case sensitive}.
\newblock   (\bibinfo{year}{2019}).
\newblock
\showURL{%
\url{https://www.zdnet.com/article/facebook-passwords-are-not-case-sensitive-update/}}


\end{thebibliography}

\end{document}